# Recognition of Surface Oxygen Intermediates on NiFe Oxyhydroxide Oxygen-evolving Catalysts by Homogeneous Oxidation Reactivity

Yaming Hao[1], Yefei Li[1*], Jianxiang Wu[1], Jinling Wang[2], Chenglin Jia[1], Tao Liu[1], Xuejing Yang[2], Zhipan Liu[1], Ming Gong[1*]

[1] Department of Chemistry and Shanghai Key Laboratory of Molecular Catalysis and Innovative Materials, Fudan University, Shanghai, China, 200438

[2]National Engineering Laboratory for Industrial Wastewater Treatment, East China University of Science and Technology, Shanghai, China, 200237

[*]Corresponding author e-mail: yefeil@fudan.edu.cn; gongm@fudan.edu.cn

## Abstract

NiFe oxyhydroxide is one of the most promising oxygen evolution reaction (OER) catalysts for renewable hydrogen production, and deciphering the identity and reactivity of the oxygen intermediates on its surface is a key challenge but is critical to understanding the OER mechanism as well as designing water-splitting catalysts with higher efficiencies. Here, we screened and utilized *in situ* reactive probes that can selectively target specific oxygen intermediates with high rates to investigate the OER intermediates and pathway on NiFe oxyhydroxide. Most importantly, the oxygen atom transfer (OAT) probes (e.g. 4-(Diphenylphosphino) benzoic acid) could efficiently inhibit the OER kinetics by scavenging the OER intermediates, exhibiting lower OER currents, larger Tafel slopes and larger kinetic isotope effect values, while probes with other reactivities demonstrated much smaller effects. Combining the OAT reactivity with electrochemical kinetic and *operando* Raman spectroscopic techniques, we identified a resting Fe=O intermediate in the Ni-O scaffold and a rate-limiting O-O chemical coupling step between a Fe=O moiety and a vicinal bridging O. DFT calculation further revealed a longer Fe=O bond formed on the surface and a large kinetic energy barrier of the O-O chemical step, corroborating the experimental results. These results point to a new direction of liberating lattice O and expediting O-O coupling for optimizing NiFe-based OER electrocatalyst.

## Introduction

Oxygen evolution reaction (OER) has played a pivotal role in the pursuit of hydrogen economy as a sustainable energy solution[1-4]. Its sluggish kinetics involving the transfer of four electrons and protons accounts for most of the energy penalty required for water electrolysis, and therefore the development of highly efficient and low-cost OER electrocatalysts is desired[3, 5-10]. In light of this, NiFe oxyhydroxide, often derived from *in situ* electrochemical oxidation of NiFe layered double hydroxide (NiFe LDH), represents one of the most active electrocatalysts for alkaline OER and has attracted tremendous attention over the past years[11-14]. The incorporation of Fe in the octahedral Ni site drastically increases the OER activity as well as reduces the Tafel slope. Its facile and scalable synthesis also makes it affordable and promising for large-scale alkaline water electrolyzers.

Recent efforts have been dedicated to understanding the molecular-level OER mechanism on NiFe oxyhydroxide surfaces to further facilitate the design and optimization of the catalyst[14-25]. A variety of electrochemical and *in situ* spectroscopic techniques have been utilized for the mechanistic studies. For instances, structural analyses with X-ray absorption spectroscopy (XAS) revealed the formation of $Fe^{4+}$ isolated sites in NiOOH as the active site with the aid of DFT calculation[16], which was further evidenced by *in situ* Mossbauer spectroscopy[17]. Scanning electrochemical microscopy (SECM) study further identified two types of sites with one fast kinetics site correlating well with the Fe content in the catalyst film, demonstrating the essential role of Fe in OER[19]. In addition to Fe, another *in situ* XAS study observed higher Ni oxidation state of +3.6 in NiFe oxyhydroxide, and attributed the OER activity to the charge transfer between Ni and Fe through Ni-O-Fe[26]. Despite that the active site is still in debate, these studies have provided important clues about the metal identities of NiFe oxyhydroxide during OER catalysis. On the other hand, the surface oxygen intermediates associated with the metal sites play indispensable roles in the redox transitions toward dioxygen generation. According to calculation studies, the optimal binding of oxygen intermediates on molecular-like sites is critical to lowering the OER energy barrier on NiFe oxyhydroxide catalysts[27-30]. However, experimental evidences are currently still insufficient and spectroscopic analyses by Raman spectroscopy and ambient pressure X-ray photoelectron spectroscopy (AP-XPS) have achieved little success in identifying the oxygen intermediates formed on NiFe oxyhydroxide[15, 22, 25]. It is largely due to the low averaged signal of transient oxygen intermediates, especially under fast turnover rates in NiFe oxyhydroxide catalysts. Therefore, developing new tools for probing oxygen intermediates under operating conditions is still highly desired for understanding the OER mechanism.

Herein, we took inspiration from the use of molecular probes in catalytic studies[31-32], and systematically designed and screened reactive probes that can interfere with OER kinetics to investigate the oxygen intermediate reactivity and the OER mechanism on NiFe oxyhydroxide. We adopted typical homogeneous oxidation reactivity of oxygen atom transfer (OAT) and hydrogen atom transfer (HAT) for the design of our probes. The probes with OAT reactivity were discovered to competitively inhibit the OER process, while probes with other reactivities exhibited smaller effects. Detailed comparison of with and without probe conditions in combination with competitive kinetics, isotope exchange experiments and *operando* Raman spectroscopy allowed us to successfully identify the key resting intermediates and the rate-determining step of OER on NiFe oxyhydroxide catalysts. According to these experimental evidences, we proposed a molecular-level OER pathway on NiFe oxyhydroxide surfaces, which was further confirmed by DFT calculation.

## Results and Discussion

## Designing and Screening of Reactive Probes

The general principle for the proposed reactive probe method utilizes the competition between the reactions of the electrochemically-generated surface oxygen intermediates with reactive probes and the OER counterparts to generate dioxygen. The introduction of reactive probes can theoretically scavenge specific oxygen intermediates and alter the electrochemical reaction pathway. Since the two pathways with and without reactive probes can drastically differ in kinetics parameters, rate-determining steps and key intermediates, we can glean the chemical properties of the oxygen intermediates as well as the molecular-level OER pathway in a comparative manner with the aid of kinetic and spectroscopic techniques (Scheme 1).

The selection of reactive probes is critical to this study. Firstly, the probe should contain a reactive center that can react with surface oxygen intermediates at high rates and high specificity. The high rates assure the efficient trapping of oxygen intermediates to compete with the OER kinetics, and the high specificity facilitates the identification of oxygen intermediates via its reactivity. We took inspiration from the well-studied reactivities in homogeneous catalytic oxidation and utilized typical reactions of oxygen atom transfer (OAT) and hydrogen atom transfer (HAT). OAT probes can specifically target oxygen-terminated structures for abstracting oxygen atoms, often including phosphines, thioethers and alkenes[33-35]. HAT probes, often C-H-based molecules, can specifically target oxygen-associated high-valent metal species, such as metal oxo and metal peroxo species, for donating H atoms to generate the corresponding radical[36-37]. These reactivities can widely exist on the oxygen intermediates formed during OER and we anticipate that probes with different reactivity can behave divergently, allowing for the differentiation and identification of oxygen intermediates. To be compatible with *in situ* experiments, the reactive probes should also demonstrate high solubility and chemical stability in the electrolyte. In addition, the probes should not contain multiple reactive centers or high degrees of freedom that may complicate the reactivity on catalyst surfaces. Accordingly, we incorporated the solubility handles of carboxylate owing to its high solubility under alkaline conditions, and the solubility handle was separated from the reactive center by a rigid aromatic scaffold (Scheme 1).

Based on the probe design principle, we selected a series of reactive probes that are either commercially available or easily accessible by a simple single-step reaction, and screened their chemical and electrochemical stability in alkaline electrolytes. For instance, the thioanisole-based probe that is typically used for investigating OAT reactions[38] exhibited considerable degradation by nucleophilic $OH^-$ attack in alkaline electrolytes, and therefore was excluded for this study (Figure S1-S3). After screening, we selected 6 representative reactive probes (2 OAT, 2 HAT and 2 control probes) (Figure S4-S15), among which we used 4-(Diphenylphosphino) benzoic acid (TPP-COOH), Fluorene-4-carboxylic acid (Fluorene-COOH) and benzoic acid (Benzene-COOH) as the OAT, HAT and control probe more extensively throughout this study.

## Competitive Kinetics Study under Probe Titration

The NiFe LDH pre-catalyst electrodes for this study were prepared by a facile electrodeposition method. The as-derived electrode exhibited uniformly-distributed particles of NiFe LDH on the carbon fibers (Figure S16). A typical CV curve involves a pair of $Ni^{2+}/Ni^{\delta+}$ ($\delta \geq 3$) redox peaks to form NiFe oxyhydroxide at 1.30-1.50 V *vs.* RHE[39], followed by the OER onset at ~1.45 V *vs.* RHE (Figure 1a). Interestingly, the titration of TPP-COOH (OAT probe) gradually lowered the OER current in a concentration-dependent manner, accompanied by the positive shift of $Ni^{2+}$ oxidation as well as the decreased current of $Ni^{\delta+}$ reduction during the reverse scan (Figure 1a). Under 5 mM TPP-COOH, all NiFe oxyhydroxide catalysts with different Fe/Ni ratios consistently exhibited ~20% OER current decay and ~30-50% $Ni^{\delta+}$ reduction current decay (Figure 1d). It indicates that the presence of surface reactions with the OAT probe greatly altered the OER kinetics. These probe effects were currently restrained by the limited solubility (~10 mM for TPP-COOH), but could be enhanced under stirring conditions to break the mass transport limitations (Figure S17). An additional OER current decay by ~11.2% and a $Ni^{\delta+}$ reduction current decay by ~19.2% could be observed on NiFe 5 catalyst (Fe/Ni=5%) at a stirring speed of 500 rpm, demonstrating the expedited reaction between the oxygen intermediates and the TPP-COOH probe. To eliminate the possibility that TPP-COOH serves as a poison to the catalyst, we carried out potentiostatic measurements with intermittent probe addition and clearly resolved instantaneous current decays upon each TPP-COOH addition (Figure S18). The current continued to decay for a certain period after the instantaneous decrease, which we attributed to the surface interaction with carboxylate that demands structural re-orientation for a maximal probe effect. Importantly, the OER activity could be almost completely recovered when reverting back to no probe conditions, excluding the irreversible poisoning effect (Figure S18).

In contrast to the OAT probe, significantly smaller changes of the OER current and $Ni^{\delta+}$ reduction current were observed upon the titration of the HAT probe (Fluorene-COOH) and the control probe (Benzene-COOH) (Figure 1b-c). The HAT probe could lead to ~10-15% OER current decay and ~15-30% $Ni^{\delta+}$ reduction current decay, while the control probe could only lead to <10% OER current decay and <15% $Ni^{\delta+}$ reduction current decay (Figure 1e-f). The general trend of OAT>HAT>control was confirmed by utilizing other probes with similar reactivities (Figure S4-S15). Since the OAT probes specifically target oxygen-terminated surfaces, the larger OAT probe effect indicates that the key OER intermediates could be oxygen-terminated species. Notably, neither OAT or HAT probes could affect the OER kinetics of the Ni oxyhydroxide catalyst without any Fe (Figure 1d-1f), suggesting that the key intermediates on Ni oxyhydroxide greatly differ from those on NiFe oxyhydroxide and follow a distinct OER pathway[40]. It also supports that the oxygen intermediates on Fe sites are the main probe target sites. By combining with the aforementioned oxygen-terminated resting state, we hypothesis that the key OER intermediates involve the molecular-like Fe=O moiety that is often subject to OAT reactivity[34-35].

Tafel analysis is a useful electrochemical tool to investigate the rate-determining step (rds) and reaction pathway during electrocatalysis. Our NiFe oxyhydroxide catalysts all exhibited typical Tafel slopes of 30-40 mV/decade, among which the NiFe 5 catalyst demonstrated the lowest Tafel slope of 31.5 mV/decade, close to a theoretical value of 30 mV/decade (Figure 1g). According to the Tafel equation and its parameter calculations (eq. 1-3)[6, 41], where $\eta$ is the OER overpotential, b is the Tafel slope, $\alpha$ is the transfer coefficient, $n_b$ is the number of electrons transferred before rds, $v$ is the number of rds in the overall reaction, $n_r$ is the number of electrons participating in the rds and $\beta$ is the symmetry factor that is often around 0.5,

$$\eta = a + b\log i \text{ (Tafel equation)} \qquad \text{eq. 1}$$

$$b = \frac{\partial \eta}{\partial \log i} = \frac{2.303RT}{\alpha F} \qquad \text{eq. 2}$$

$$\alpha = \frac{n_b}{v} + n_r\beta \qquad \text{eq. 3}$$

a Tafel slope of 30 mV/decade implies a possible rate-limiting chemical step and two electron transfer before this chemical step with $n_b$=2, $v$=1 and $n_r$=0[42]. Upon the titration of TPP-COOH, the Tafel slope showed a step-wise increase from ~30 mV/decade to ~40 mV/decade (Figure 1g). The high-rate reaction between the resting oxygen intermediate and TPP-COOH probe causes the initial rds to be no longer rate-limiting and shifts the rds toward prior steps with the second highest energy barrier. In accordance, a Tafel slope of 40 mV/decade corresponds to the rds of the second electron-transfer or PCET (with $n_b$=1, $v$=1 and $n_r$=1), which could be one step prior to the original rds with 30 mV/decade.

In previous reports, two possible rds with high energy barriers were proposed: one is the oxidation step of Fe-OH to Fe=O, and the other is the O-O coupling step[27-30]. Based on our experiments, the rate-limiting oxidation of Fe-OH to Fe=O is very unlikely due to the reasons that 1) transferring two electrons before the rds can hardly be achieved, 2) the resting state of Fe-OH cannot be kinetically affected by OAT probes and 3) the rds of Fe-OH oxidation is not a chemical step that is inconsistent with a Tafel slope of 30 mV/decade. The rate-limiting O-O coupling step was relatively a better match[29]. The Fe=O resting state has strong OAT reactivity and can be efficiently scavenged by the OAT probe to alter the kinetics. Meanwhile, the catalytic cycle upon probe addition shifts the O-O coupling rds to the former Fe-OH oxidation step that is highly consistent with the Tafel study. In contrary to the OAT probes, the Benzene-COOH control probe demonstrated negligible changes to Tafel slopes under varied doses (Figure 1i), while the Fluorene-COOH HAT probe could reproducibly elevate the Tafel slope by a minor extent of 2-3 mV/decade (Figure 1h). This effect could be expected due to the typical HAT reactivity of the proposed Fe=O intermediate that is often found in heme or non-heme catalysts[36]. Since HAT reactivity highly depends on the C-H bond strength, the subtle differences disappeared when using 4-methylbenzoic acid (Toluene-COOH) as the probe, suggesting the incapability of benzylic C-H oxidation by the oxygen intermediates on the NiFe oxyhydroxide catalyst (Figure S10-11). Despite that Fluorene-COOH has one of the weakest C-H bonds, it still possesses tremendously lower effects than the OAT probe, which indicates that the electrochemically generated Fe=O intermediates on hydroxide/oxyhydroxide surfaces has a larger tendency to donate the O atom with OAT reactivity than abstract an H atom with HAT reactivity. This reactivity greatly differs from the Fe-based homogeneous catalyst, and further points to the uniqueness of the hydroxide/oxyhydroxide scaffold for OER catalysis.

One of the unique advantages of using reactive probes is the high versatility of combining with other techniques for the comprehensive understanding of the elemental reaction steps. We further incorporated the probe strategy into the kinetic isotope H/D exchange experiments by comparing 1 M KOD in $D_2O$ and 1 M KOH in $H_2O$ electrolytes side by side to investigate the participation of H in the rds (Figure 2a-b and Figure S19)[32]. Without any probe, the catalyst exhibited parallel Tafel curves but with slightly inferior OER current in KOD to that in KOH. The $k_H/k_D$ value was further calculated to be in the range of 1.37-1.40, representing a secondary kinetic isotope effect (KIE) that does not involve direct H-related bond breaking process in the rds (Figure 2b). In addition to the OER current decrease and Tafel slope increase, increasing doses of TPP-COOH displayed a steady increase of $k_H/k_D$ values, reaching 1.64-1.67 at 7.5 mM TPP-COOH (Figure 2b). The increasing KIE values suggest the shift of a secondary KIE toward a primary KIE that involves direct H-related bond breaking event as the limiting step. These results agree well with our proposal of the rate-determining O-O coupling step with no direct H participation and the rate-limiting O-H bond breaking during Fe-OH oxidation into Fe=O after probe addition.

In spite of the identification of the resting Fe=O intermediates, the molecular aspects of the rate-limiting O-O step are less understood. When analyzing the $Ni^{\delta+}$ reduction and OER current decay upon probe addition, we obtained an interesting Pearson correlation coefficient of 0.867, indicating the strong correlation between the two parameters (Figure S20). Since the structural Fe sites are mainly accounted for the probe reaction as indicated earlier, such correlation might represent the strong electronic communication between Ni and Fe sites, or in other words, the Ni sites may participate in the rate-limiting O-O step despite that the key oxygen intermediates are located on Fe sites. Based on the Fe=O intermediate and rate-limiting chemical step, there might be two possible origins of the O-O coupling step: 1) the formation of surface-bound OOH by the transfer of one solution-phase $OH^-$ onto Fe=O; 2) direct O-O coupling of Fe=O by the lattice O or OH. These two possibilities could give rise to different Nernst potential shifts per pH, since the involvement of additional $OH^-$ in the rds could lead to a larger Nernst shift. Under different pHs in the 12.5-14.0 regime, we observed parallel Nernst shifts of the polarization curves (Figure 2c). By plotting the potentials at a current density of 1 $mA/cm^2$ over pHs (this current density was selected to minimize the influence of different Tafel slopes), we could derive the potential shifts of 61.1 mV/pH and 62.2 mV/pH without and with the addition of TPP-COOH respectively, which are very close to the theoretical value of 59.1 mV/pH that corresponds to one $H^+/OH^-$ per electron transfer (Figure 2c-d)[6]. We attribute this pH shift to a rate-limiting lattice O-driven O-O coupling rather than the attack of solution-phase $OH^-$ during the rds, because otherwise it should involve total of 3 $OH^-$ and $2e^-$ transfer corresponding to a larger potential shift of ~90 mV/pH accordingly[43]. Under TPP-COOH, the shift of rds toward Fe-OH oxidation involves a total of 2 $OH^-$ and $2e^-$ transfers, consistent with the observed Nernst shift. Since the lattice O represents the bridging O between Fe and Ni, its participation in OER could explain the strong electronic communication between Ni and Fe sites in the catalyst.

## *Operando* Raman Spectroscopy with OAT Probes

We further combined our reactive probe approach with spectroscopic techniques to facilitate the identification of surface oxygen intermediates. We first confirmed the OAT and HAT reactivities by the mass spectroscopy (MS) and nuclear magnetic resonance (NMR) analysis of the transformation products after potentiostatic electrolysis (Figure S21-22). Complete conversion of TPP-COOH into TPPO-COOH under typical OAT reactivity was observed, while only partial oxidation of Fluorene-COOH was detected under identical conditions. It is in consensus with the observed trend in the probe effects. Owing to its compatibility with aqueous electrolyte system for *operando* measurements, Raman spectroscopy has been widely adopted to probe the reactive intermediates during catalytic OER processes. For examples, Bell et al. identified a surface bound OOH intermediate on gold electrodes by Raman spectroscopy[44]. Hu et al. combined the Raman spectroscopy with the active NiFe-based electrocatalysts to correlate the OER activity with structural disorder[15]. We envision that by using reactive probes such as TPP-COOH, an increase or decrease of the intermediate signal should be viable due to the efficient trapping of the oxygen intermediates. The Raman spectra showed characteristic peaks of $\delta$ ($Ni^{III}$-O) and $\upsilon$ ($Ni^{III}$-O) band at Raman shifts of 474 $cm^{-1}$ and 544 $cm^{-1}$ respectively as well as the broad bands at around 800-1200 $cm^{-1}$ previously attributed to the superoxide anion ($-OO^{-}$) intermediates, which are consistent with previous studies[14] (Figure 3b and S23). Astonishingly, we observed highly resembling Raman patterns upon TPP-COOH addition except for the appearance of a peak at 1066 $cm^{-1}$ (Figure 3b). By analyzing the liquid electrolyte before and after Raman measurement, we observed the evolution of the 1066 $cm^{-1}$ peak in addition to the aromatic peak at 1000 $cm^{-1}$, which was possibly attributed to the P=O bond in the oxidized probe (Figure 3a). Its successful detection by *operando* Raman spectroscopy not only confirmed the OAT reactivity but also support the scavenge of oxygen-terminated species upon the addition of TPP-COOH. Although negligible changes were observed in addition to the P=O peak (Figure S24), the identical patterns at 800-1200 $cm^{-1}$ might evidence that the broad peak of superoxide might not be related to OER, because TPP-COOH should hinder the OER pathway and bypass superoxide formation, theoretically leading to a significant decrease in the superoxide signal. This speculation correlates well with the presence of this band under non-OER conditions. We render that this band may belong to the bridging superoxide species or other surface oxygen species that is structurally relevant to $Ni^{2+}$ oxidation but less relevant to OER.

Despite the overall similarity of the Raman spectra, we focused our attention to the Ni-O peak region of 400-650 $cm^{-1}$ and discovered subtle changes under different potentials with the addition of TPP-COOH. Specifically, we normalized the Raman spectra based on the 474 $cm^{-1}$ peak and discovered that the 554 $cm^{-1}$ peak was slightly narrower under TPP-COOH (Figure 3c-d). The peak width at 80% intensity was narrower by 4-6 $cm^{-1}$ and relatively peak integration ratio of $I_{554}/I_{474}$ was lower by 10-15% under OER conditions compared to non-OER conditions (Figure 3e-f). Such differences were only observed with the presence of TPP-COOH, advocating the significance of using reactive probes for *operando* spectroscopic study. We calculated the theoretical Raman shifts of terminal Fe=O and bridging Fe-O-Ni species to be around 639 $cm^{-1}$ and 531, 586 $cm^{-1}$ respectively (Figure S25), and therefore the narrower peak at 554 $cm^{-1}$ was unlikely caused by the diminishing Fe=O intermediates, but instead was derived from bridging O species near the Fe centers. In previous reports, the relative ratio of $I_{554}/I_{474}$ is a signature of the structural disorder in Ni oxyhydroxide, and the incorporation of Fe can induce the elevation of this value by increasing structural disorder, which correlates well with the OER activity[22]. According to our observations, we speculated that the efficient scavenge of Fe=O intermediates on the surfaces by OAT probes can reduce the Fe centers and decrease the local structural disorder of the Ni scaffold by maintaining the structure at a relatively low oxidation state, which in turn decelerated the OER process. This decrease in local structural disorder was possibly stemmed from the vanishing bridging O species in Fe-O-Ni and the decreased availability of bridging O species, affect the OER kinetics due to its strong involvement in the rate-determining O-O coupling step as lattice O. To our knowledge, this could serve as the first spectroscopic evidence of the participation of bridging O species in Fe-O-Ni during the OER of NiFe oxyhydroxide catalyst. As Fe=O intermediates may have relatively fast turnover rates and low surface densities, and Raman studies of Fe=O intermediates in heme-based catalysts often require freeze-quench techniques, we could only obscure peaks for the differential Raman spectra in the Raman shift range of 600-700 $cm^{-1}$ that might belong to the diminishing Fe=O intermediates under highly oxidative potentials (Figure S26). Therefore, further efforts on spectroscopically identifying Fe=O intermediates on the NiFe oxyhydroxide surfaces are still desired and undergoing.

## OER mechanism and Theoretical Calculation

Based on our kinetic and spectroscopic experiments, we carried out the systematic deduction of the molecular-level OER pathway (Figure 4a). The initial state is critical to the mechanistic analysis, because the kinetic parameters can vary drastically according to the different initial states. Since OER is only observed after the completion of $Ni^{2+}$ oxidation and Fe is mostly stable in its +3 oxidation state before the OER potential, we anticipate that the initial state involves +3 Ni and +3 Fe, which can be viewed as the one-electron oxidized form of NiFe LDH with +2 Ni and +3 Fe. Without applied potential, the central site of NiFe LDH can be abbreviated as $*FeOH(\mu_2\text{-}OH)_2$ with two $\mu_2$-OHs connected to the +2 Ni; with applied OER potential, one-electron oxidation of Ni is accompanied by the release of one proton into the solution, generating $*FeOH(\mu_2\text{-}O)(\mu_2\text{-}OH)$ with the $\mu_2$-O connected to +3 Ni. This $*FeOH(\mu_2\text{-}O)(\mu_2\text{-}OH)$ state is designated as the initial state for OER.  Since the transfer of two electrons has to be completed before the rds according to the Tafel analysis, the initial state undergoes consecutive electron transfers from the initial state of +3 Ni / +3 Fe to form the resting state containing either +4 Ni / +4 Fe or +3 Ni / +5 Fe. As +4 Ni is thermodynamically more favorable than +5 Fe and higher Ni oxidation state was previously detected upon incorporation of Fe into the Ni-based catalyst[26], the key intermediates are more likely involving +4 Ni / +4 Fe, denoted as $*FeO(\mu_2\text{-}O)_2$ with two $\mu_2$-Os connected to +4 Ni. In this $*FeO(\mu_2\text{-}O)_2$ intermediate, Fe=O is mainly responsible for the greatly affected OER kinetics by the OAT reactivity. Following the resting intermediate, the rds is the subsequent step of breaking Ni-O bond and transfer of O to form a cyclic Fe peroxide intermediate with O-O coupling as its chemical nature and the product denoted as $*Fe(\mu_2\text{-}O)(\eta^2\text{-}O_2)$. This step is a chemical step, which is consistent with the Tafel slope and kinetic isotope experiments, but from kinetic aspects, it could be rather unfavored by involving one Ni-O bond breaking and one O-O bond formation in a single transition. As a consequence, it might involve large chemical energy barriers to be rate-limiting to the entire pathway. The use of lattice O or $\mu_2$-O for dioxygen formation is evidenced by the Nernst shift and Raman spectroscopy study. Following the rds, the as-derived cyclic peroxide intermediate can further release the oxygen and eventually be subject to a series of $OH^-$ and electron transfers to regenerate the surface. Under the addition of TPP-COOH, the resting state of $*FeO(\mu_2\text{-}O)_2$ is rapidly scavenged by OAT reactivity to form $*Fe(\mu_2\text{-}O)_2$, which can be subject to a series of reactions to regenerate the initial state but bypassing the OER pathway.

In order to further validate our proposed mechanism, we carried out the density functional theory (DFT) calculations for the molecular-level OER pathway (see Supplementary Information for the calculation details). We used a deprotonating β-NiOOH structure by removing 2/3 protons to model the structure of γ-NiOOH and specifically used the γ-NiOOH $(\bar{1}2\bar{1}0)$ surface as the model surface (Figure S27). Our previous study on γ-NiOOH has shown that the calculations based on this simplified model can yield the overpotential comparable with the experimental data[45]. We used a potential of U=1.63 V vs. RHE to represent the pathway under OER conditions. Based on the initial state of $FeOH(\mu_2\text{-}O)(\mu_2\text{-}OH)$, the first PCET step facilely releases a proton from the surface upon oxidation ($*FeOH(\mu_2\text{-}O)(\mu_2\text{-}OH) \rightarrow *FeOH(\mu_2\text{-}O)_2 + H^+ + e^-$) with a Gibbs free energy change of +0.03 eV (Figure 4b-c). After that, the terminal *Fe-OH can undergo a second PCET to form *Fe=O, ($*FeOH(\mu_2\text{-}O)_2 \rightarrow *FeO(\mu_2\text{-}O)_2 + H^+ + e^-$), and this step has a relatively higher Gibbs free energy change of 0.51 eV. Interestingly, the bond length of Fe=O is determined to be 1.74 Å, which is longer than that in conventional iron-oxo compounds. The elongation of the Fe=O bond can favor the OAT reaction by donating the O atom to the substrate, advocating the strong influence of OAT probe to the OER kinetics and implying the probably higher tendency of forming O-O bonds on this Fe=O site. Despite that the following step of ($*FeO(\mu_2\text{-}O)_2 \rightarrow *Fe(\mu_2\text{-}O)(\eta^2\text{-}O_2)$) is exothermal by 1.03 eV, this chemical step has a relatively high energy barrier of 0.41 eV, comparable to the former step (Figure 4c). It is worthwhile to mention that the nature of a chemical step determines that its energy barrier is significantly less influenced by the potentials than other electron-involved steps. Consequently, this energy barrier of the chemical step can be a rate-determining factor for the OER process, which is highly consistent with the experimental results. After the O-O coupling, an $O_2$ molecule desorbs from the $\eta^2$-$O_2$ site, and then water molecules heal the remaining lattice O vacancy to regenerate the initial state *FeOH($\mu_2$-O)($\mu_2$-OH). These steps are highly exothermic and require less energy input comparing with the first three steps, so we skip the details of these processes. Moreover, we also calculated the free energy change of the OAT reaction between the active *FeO($\mu_2$-O)$_2$ and the reactive TPP core (*FeO($\mu_2$-O)$_2$ + TPP → *FeO-TPP($\mu_2$-O)$_2$) to demonstrate the high rate OAT reactivity. The result turns out that this reaction is strongly exothermic with a free energy change of -4.56 eV, suggesting almost spontaneous oxygen abstraction upon TPP-based probe addition. This high-rate reaction sets up the basis for the mechanistic study by the probe approach. This mechanism is enabled only by combining the kinetic and spectroscopic study with the reactive probe approach, demonstrating a practical way of investigating mechanistic pathways with high turnover rates and transient intermediates using probes.

## Conclusion

In conclusion, we have screened and utilized the reactive probes with homogeneous oxidation reactivity to investigate the chemo-reactivity of oxygen intermediates during OER on NiFe oxyhydroxide catalysts. The representative oxygen atom transfer (OAT) probe of 4-(Diphenylphosphino) benzoic acid (TPP-COOH)) can effectively react with oxygen intermediates and alter the OER kinetics by lowering ~20% OER current and increasing the Tafel slope by up to 10 mV/decade. Comparatively, the hydrogen atom transfer (HAT) probes and control probes exhibited significantly lower effects. The strong influence on OER by OAT reactivity manifests the key intermediates of oxygen-terminated surfaces. Further combination of the negligible probe effect on Ni oxyhydroxide catalysts and the altered kinetic isotope effect (KIE) values under TPP-COOH successfully identifies the resting Fe=O intermediate during OER. The calculated Fe=O bond length of ~1.74 Å is much larger than conventional Fe=O bonds, corroborating the favored OAT reactivity. Moreover, incorporating the OAT probe into *operando* Raman spectroscopy reveals the declination of bridging O signals in Fe-O-Ni moieties, suggesting the possible participation of bridging O in the rate-determining step (rds). Further evidences by Nernst shift and Tafel analysis under probe conditions point to the rds of O-O coupling between Fe=O and a nearby bridging O. According to these findings, we propose a molecular-level OER mechanism on NiFe oxyhydroxide and DFT calculation confirms its feasibility by demonstrating a relatively high energy barrier of the chemical O-O coupling step. Expediting O-O coupling and liberating lattice O is, thus, highly critical for NiFe catalyst design to further increase the catalytic efficiency. This mechanism was only enabled by the combination of reactive probes and kinetic/spectroscopic techniques, which can be further generalized to other catalyst systems.

## Acknowledgement

M. Gong acknowledges supports from the National Key R&D program of China (2019YFC1604602). X. J. Yang acknowledges supports from the National Key Basic Research Program of China (2019YFC1906700) and National Natural Science Foundation of China (21876049, 51878643). Y. F. Li acknowledges supports from National Natural Science Foundation of China (21972023, 21773032).

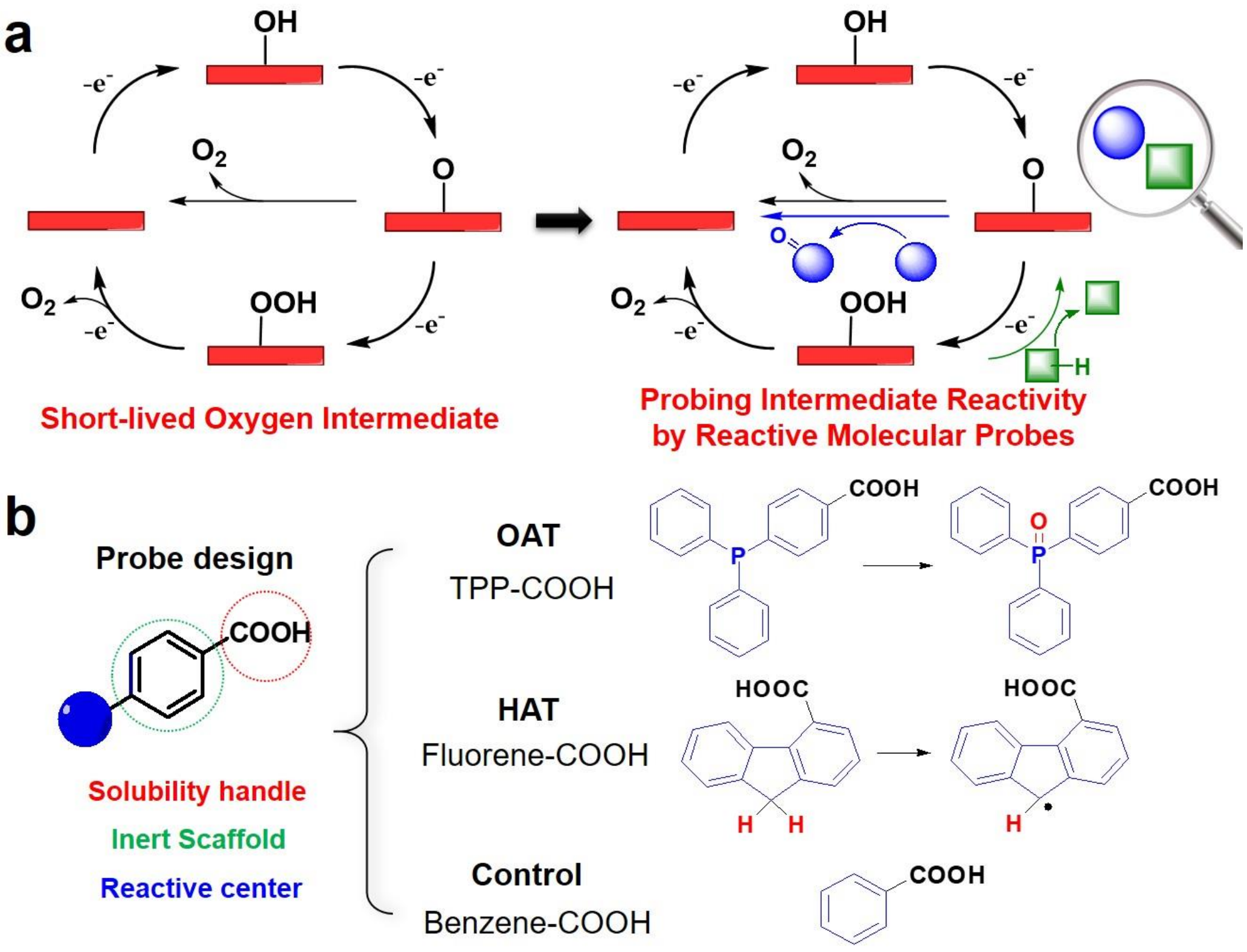

**Scheme 1**. a) Schematic illustration of the probe method for the OER mechanistic study. The high-rate reactivity of the probe with the oxygen intermediates can tackle the challenging problem of transient intermediates that restricts current mechanistic study. b) The design principle of the molecular probes and the representative probes used in the current study.

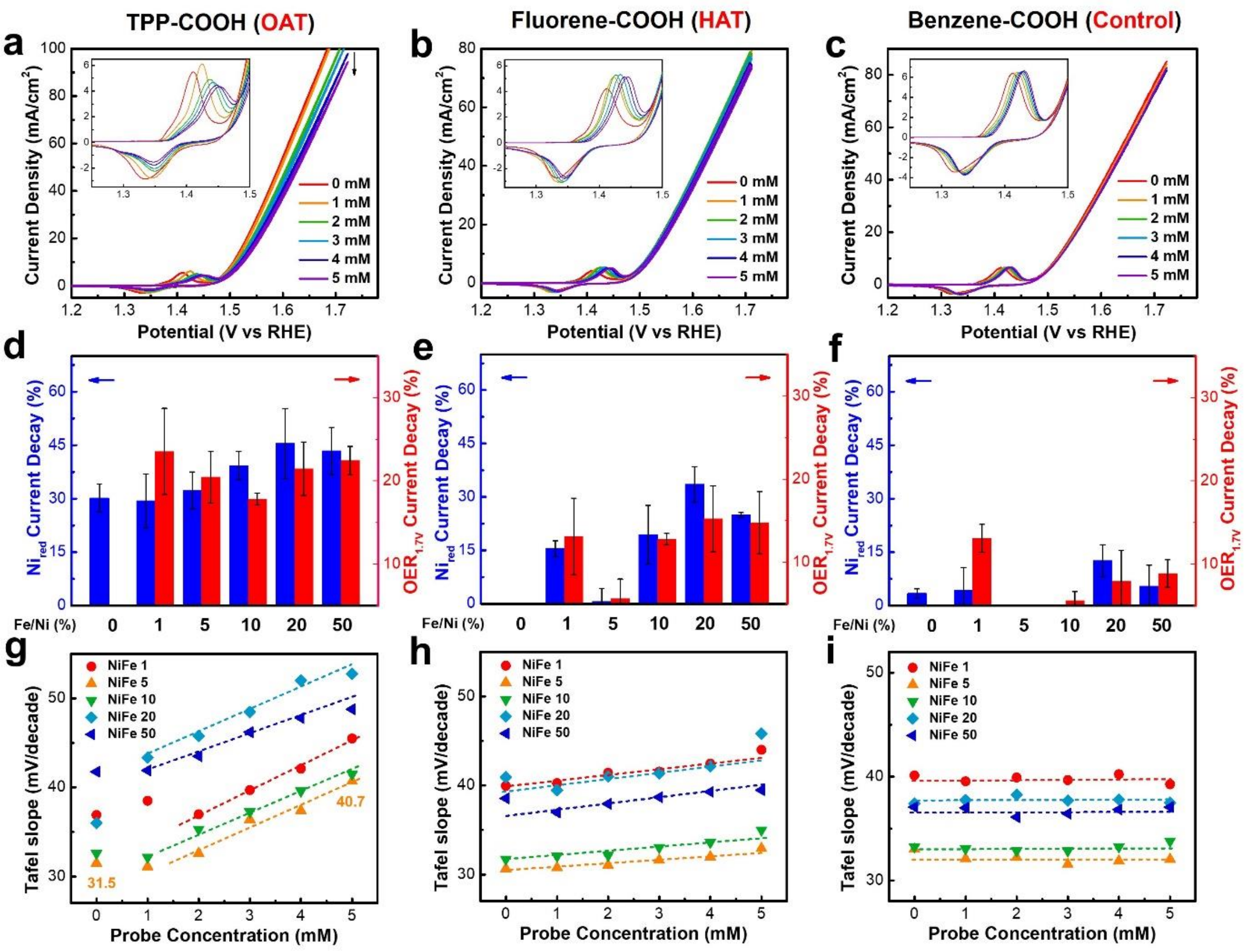

**Figure 1**. a-c) Cyclic voltammetry curves of the NiFe 5 (Fe/Ni=5%) catalysts in 1 M KOH with the titration of 0-5 mM probes, a) TPP-COOH, b) Fluorene-COOH, c) Benzene-COOH. The scan rate is 5 mV/s. The insets show the Ni redox features upon the probe titration. d-f) The change in the two parameters of $Ni^{\delta+}$ reduction current and OER current at 1.7 V vs RHE over different NiFe LDH catalysts with varied Fe/Ni ratios under the addition of 5 mM d) TPP-COOH, e) Fluorene-COOH, f) Benzene-COOH in the electrolyte. The parameters are derived from the cyclic voltammetry curves, which are listed in Figure S1-S12. g-i) The calculated Tafel slopes over different NiFe LDH catalysts with varied Fe/Ni ratios under different probe concentrations of g) TPP-COOH, h) Fluorene-COOH, i) Benzene-COOH. The resistances are reported in the Supplementary Information

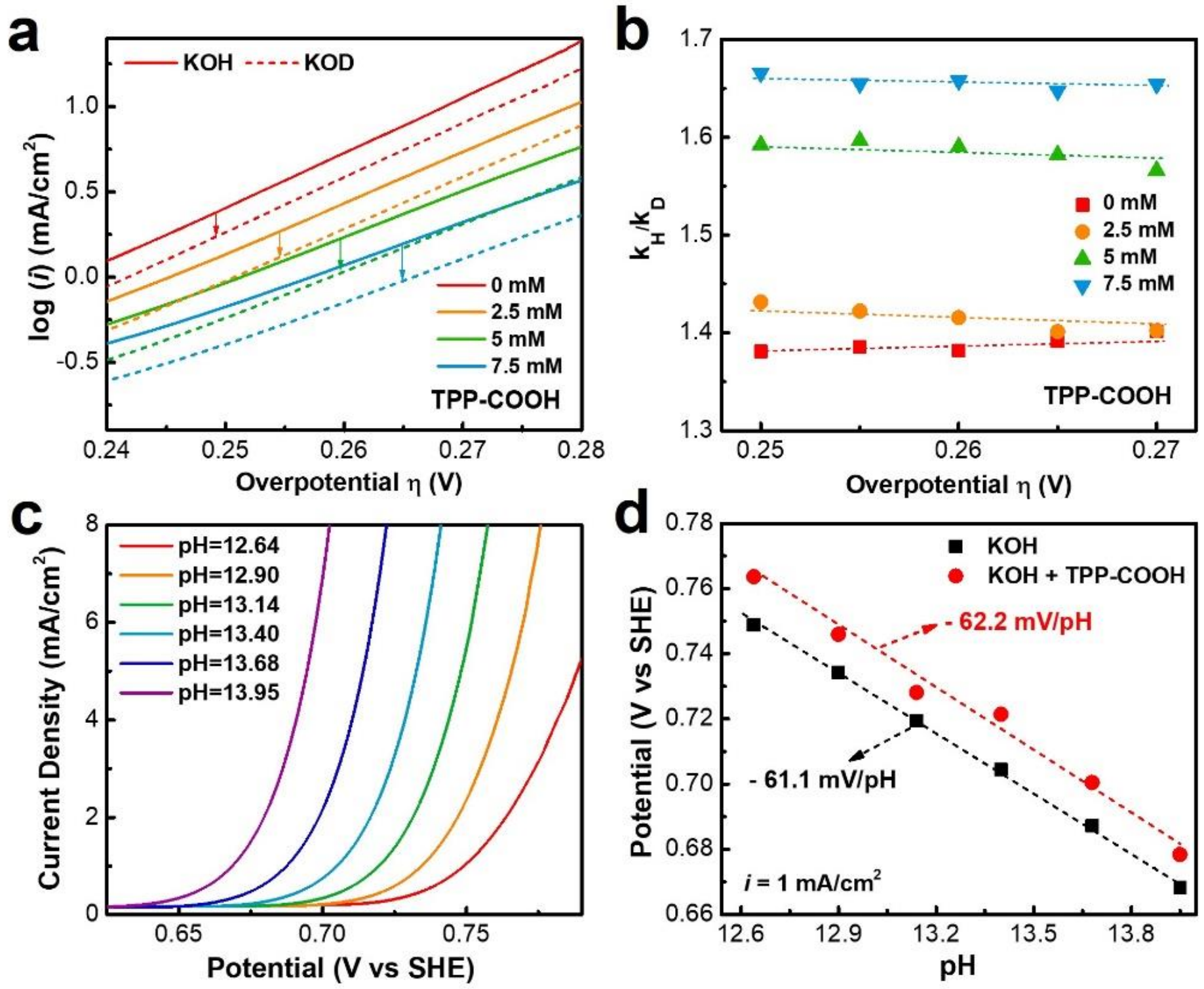

**Figure 2**. a-b) The Tafel plots of NiFe 5 (Fe/Ni=5%) catalysts in 1 M KOH with varied concentrations of TPP-COOH under H/D exchange (1 M KOH in $H_2O$ and 1 M KOD in $D_2O$). The scan rate is 0.5 mV/s b) The calculated $k_H/k_D$ values under varied concentrations of TPP-COOH over different overpotentials. c) The typical cyclic voltammetry curves of NiFe 5 catalysts under different pHs. (R=1.8Ω) d) The potentials at current densities of 1 mA/cm$^2$ on NiFe 5 catalysts in 1 M KOH and 1 M KOH + 5 mM TPP-COOH under different pHs. The slope corresponds to the Nernst shift.

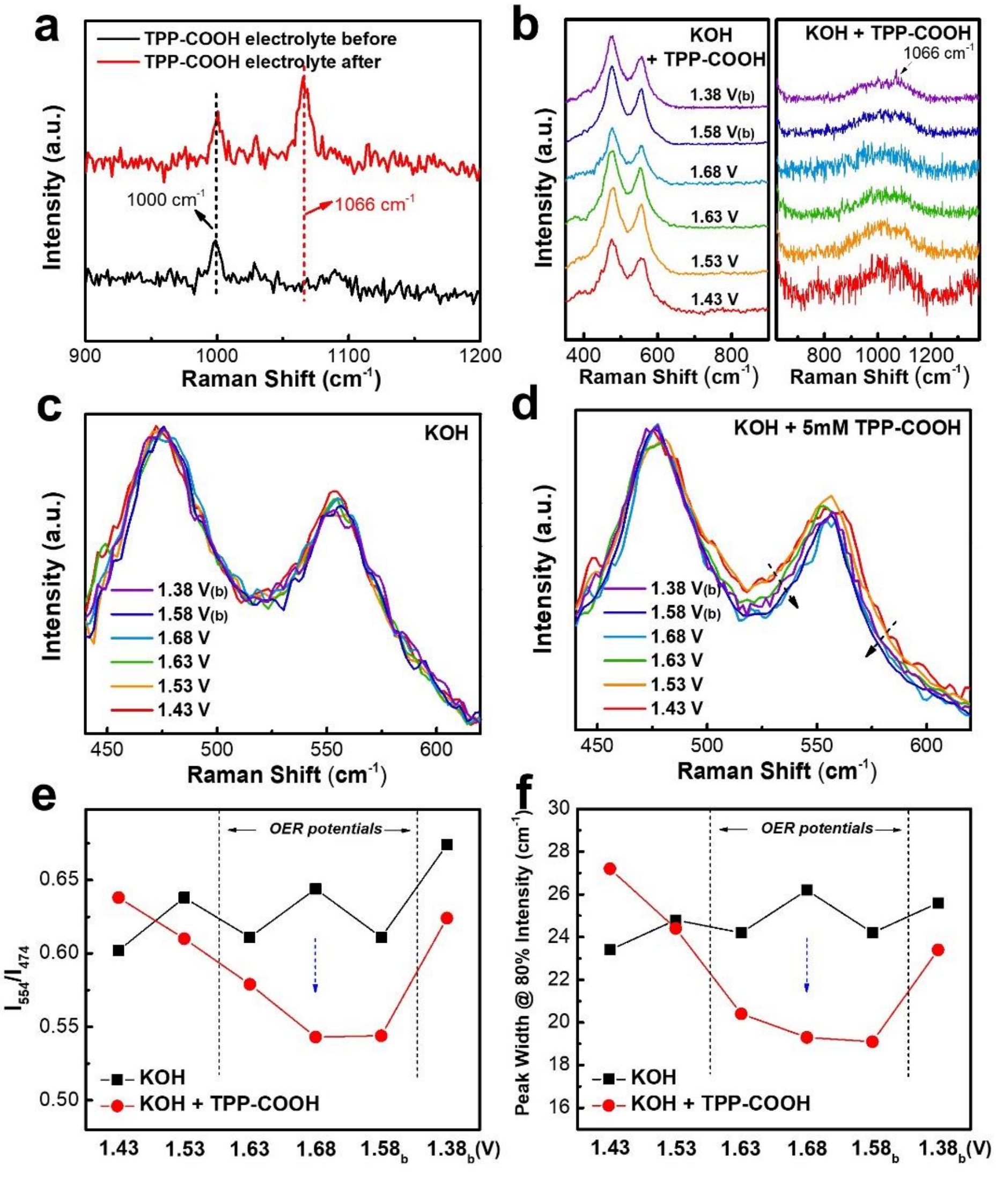

**Figure 3**. a) the Raman spectra of the electrolyte (1 M KOH + 5 mM TPP-COOH) before and after electrolysis for the Raman study. b) The Raman spectra of the NiFe 5 catalyst in 1 M KOH + 5 mM TPP-COOH under different applied potentials. c-d) Comparison of the normalized Raman spectra of the NiFe 5 catalyst in c) 1 M KOH and d) 1 M KOH + 5 mM TPP-COOH under different applied potentials. e-f) the calculated relative intensity ratio of $I_{554}/I_{474}$ and the 554 $cm^{-1}$ peak width at 80% peak intensity in 1 M KOH and 1 M KOH + 5 mM TPP-COOH under different applied potentials.

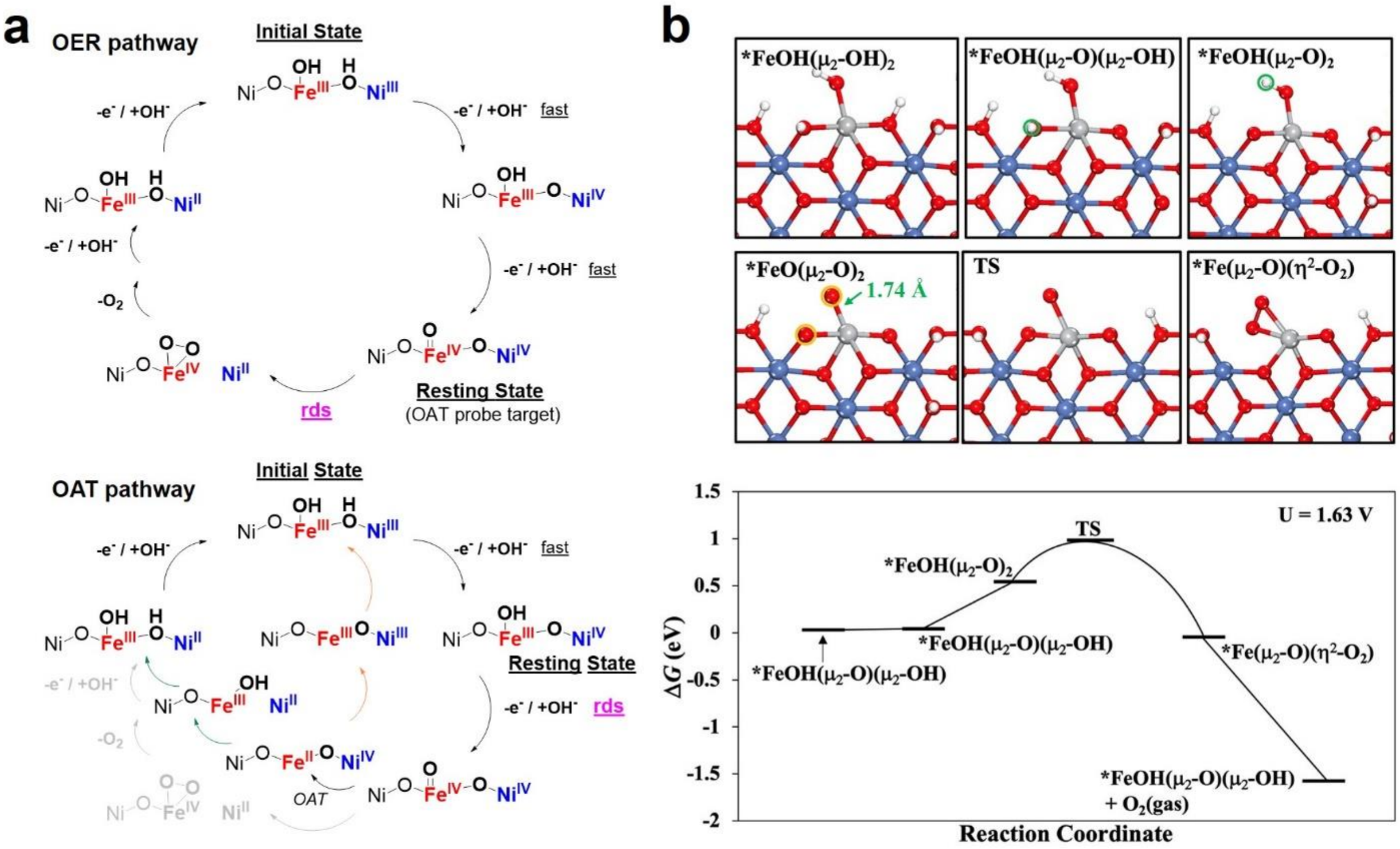

**Figure 4**. a) The proposed mechanism of OER catalytic cycle with and without the OAT probe addition. b) Structures for key intermediate states of OER and energetic profiles for OER on Fe-doped γ-NiOOH $(\bar{1}2\bar{1}0)$. Free energy changes are calculated at the electrode potential (U) at 1.63 V vs. RHE. The small green cycles in (a) high light the protons that release in the subsequent step. The small green cycles highlight the O atoms that generate a surface peroxide. *FeOH($\mu_2$-OH)$_2$ is the surface state without applied potential, while *FeOH($\mu_2$-O)($\mu_2$-OH) is the initial state for OER. The nomenclature of coordination complexes (e.g., μ and η) is utilized to represent the terminal structures of Fe ion. The asterisk * represents the Fe ion is a surface atom of γ-NiOOH. Blue balls: Ni; gray balls: Fe; red balls: O; white balls: H.

# References

1. Katsounaros, I.; Cherevko, S.; Zeradjanin, A. R.; Mayrhofer, K. J., Oxygen electrochemistry as a cornerstone for sustainable energy conversion. *Angew Chem Int Ed Engl* **2014,** *53* (1), 102-21.
2. Gray; Harry, B., Powering the planet with solar fuel. *Nature Chemistry* **2009,** *1* (1), 7.
3. Carmo, M.; Fritz, D. L.; Mergel, J.; Stolten, D., A comprehensive review on PEM water electrolysis. *International Journal of Hydrogen Energy* **2013,** *38* (12), 4901-4934.
4. Zeng, K.; Zhang, D., Recent progress in alkaline water electrolysis for hydrogen production and applications. *Progress in Energy and Combustion ence* **2010,** *36* (3), 307-326.
5. Darcy, J. W.; Koronkiewicz, B.; Parada, G. A.; Mayer, J. M., A Continuum of Proton-Coupled Electron Transfer Reactivity. *Acc Chem Res* **2018,** *51* (10), 2391-2399.
6. Suen, N. T.; Hung, S. F.; Quan, Q.; Zhang, N.; Xu, Y. J.; Chen, H. M., Electrocatalysis for the oxygen evolution reaction: recent development and future perspectives. *Chem. Soc. Rev.* **2017,** *46* (2), 337.
7. Li, J.; Güttinger, R.; Moré, R.; Song, F.; Wan, W.; Patzke, G. R., Frontiers of water oxidation: the quest for true catalysts. *Chem. Soc. Rev.* **2017,** *46* (20), 6124-6147.
8. Jiao, Y.; Zheng, Y.; Jaroniec, M.; Qiao, S. Z., Design of electrocatalysts for oxygen- and hydrogen-involving energy conversion reactions. *Chem. Soc. Rev.* **2015,** *46* (8), 2060-2086.
9. Wang, Y.; Yan, D.; El Hankari, S.; Zou, Y.; Wang, S., Recent Progress on Layered Double Hydroxides and Their Derivatives for Electrocatalytic Water Splitting. *Advanced Science* **2018,** *5* (8), 1800064.
10. Lv, L.; Yang, Z.; Chen, K.; Wang, C.; Xiong, Y., 2D Layered Double Hydroxides for Oxygen Evolution Reaction: From Fundamental Design to Application. *Advanced Energy Materials* **2019,** *9* (17), 1803358.
11. Gong, M.; Li, Y.; Wang, H.; Liang, Y.; Wu, J. Z.; Zhou, J.; Wang, J.; Regier, T.; Wei, F.; Dai, H., An Advanced Ni-Fe Layered Double Hydroxide Electrocatalyst for Water Oxidation. *Journal of the American Chemical Society* **2013,** *135* (23), 8452-8455.
12. Gong, M.; Dai, H., A mini review of NiFe-based materials as highly active oxygen evolution reaction electrocatalysts. *Nano Research* **2015,** *8* (1), 23-39.
13. Dionigi, F.; Strasser, P., NiFe-Based (Oxy)hydroxide Catalysts for Oxygen Evolution Reaction in Non-Acidic Electrolytes. *Advanced Energy Materials* **2016,** *6* (23), 1600621.
14. Trzesniewski, B. J.; Diaz-Morales, O.; Vermaas, D. A.; Longo, A.; Bras, W.; Koper, M. T.; Smith, W. A., In Situ Observation of Active Oxygen Species in Fe-Containing Ni-Based Oxygen Evolution Catalysts: The Effect of pH on Electrochemical Activity. *J Am Chem Soc* **2015,** *137* (48), 15112-21.
15. Lee, S.; Banjac, K.; Lingenfelder, M.; Hu, X., Oxygen Isotope Labeling Experiments Reveal Different Reaction Sites for the Oxygen Evolution Reaction on Nickel and Nickel Iron Oxides. *Angewandte Chemie International Edition* **2019,** *58* (30), 10295-10299.
16. Friebel, D.; Louie, M. W.; Bajdich, M.; Sanwald, K. E.; Cai, Y.; Wise, A. M.; Cheng, M.-J.; Sokaras, D.; Weng, T.-C.; Alonso-Mori, R.; Davis, R. C.; Bargar, J. R.; Norskov, J. K.; Nilsson, A.; Bell, A. T., Identification of Highly Active Fe Sites in (Ni,Fe)OOH for Electrocatalytic Water Splitting. *Journal of the American Chemical Society* **2015,** *137* (3), 1305-1313.
17. Chen, J. Y. C.; Dang, L.; Liang, H.; Bi, W.; Gerken, J. B.; Jin, S.; Alp, E. E.; Stahl, S. S., Operando Analysis of NiFe and Fe Oxyhydroxide Electrocatalysts for Water Oxidation: Detection of Fe4+ by Mössbauer Spectroscopy. *Journal of the American Chemical Society* **2015,** *137* (48), 15090-15093.
18. Louie, M. W.; Bell, A. T., An Investigation of Thin-Film Ni-Fe Oxide Catalysts for the Electrochemical Evolution of Oxygen. *Journal of the American Chemical Society* **2013,** *135* (33), 12329-12337.
19. Ahn, H. S.; Bard, A. J., Surface Interrogation Scanning Electrochemical Microscopy of Ni(1-x)Fe(x)OOH ($0 < x < 0.27$) Oxygen Evolving Catalyst: Kinetics of the "fast" Iron Sites. *J Am Chem Soc* **2016,** *138* (1), 313-8.

20. Zhang, J.; Liu, J.; Xi, L.; Yu, Y.; Chen, N.; Sun, S.; Wang, W.; Lange, K. M.; Zhang, B., Single-Atom Au/NiFe Layered Double Hydroxide Electrocatalyst: Probing the Origin of Activity for Oxygen Evolution Reaction. *Journal of the American Chemical Society* **2018,** *140* (11), 3876-3879.
21. Zhu, K.; Zhu, X.; Yang, W., Application of In Situ Techniques for the Characterization of NiFe-Based Oxygen Evolution Reaction (OER) Electrocatalysts. *Angewandte Chemie International Edition* **2019,** *58* (5), 1252-1265.
22. Lee, S.; Bai, L.; Hu, X., Deciphering Iron-Dependent Activity in Oxygen Evolution Catalyzed by Nickel–Iron Layered Double Hydroxide. *Angewandte Chemie International Edition* **2020,** *59* (21), 8072-8077.
23. Sayler, R. I.; Hunter, B. M.; Fu, W.; Gray, H. B.; Britt, R. D., EPR Spectroscopy of Iron- and Nickel-Doped [ZnAl]-Layered Double Hydroxides: Modeling Active Sites in Heterogeneous Water Oxidation Catalysts. *Journal of the American Chemical Society* **2020,** *142* (4), 1838-1845.
24. González-Flores, D.; Klingan, K.; Chernev, P.; Loos, S.; Mohammadi, M. R.; Pasquini, C.; Kubella, P.; Zaharieva, I.; Smith, R. D. L.; Dau, H., Nickel-iron catalysts for electrochemical water oxidation – redox synergism investigated by in situ X-ray spectroscopy with millisecond time resolution. *Sustainable Energy & Fuels* **2018,** *2* (9), 1986-1994.
25. Ali-Löytty, H.; Louie, M. W.; Singh, M. R.; Li, L.; Sanchez Casalongue, H. G.; Ogasawara, H.; Crumlin, E. J.; Liu, Z.; Bell, A. T.; Nilsson, A.; Friebel, D., Ambient-Pressure XPS Study of a Ni–Fe Electrocatalyst for the Oxygen Evolution Reaction. *The Journal of Physical Chemistry C* **2016,** *120* (4), 2247-2253.
26. Li, N.; Bediako, D. K.; Hadt, R. G.; Hayes, D.; Kempa, T. J.; von Cube, F.; Bell, D. C.; Chen, L. X.; Nocera, D. G., Influence of iron doping on tetravalent nickel content in catalytic oxygen evolving films. *Proceedings of the National Academy of Sciences* **2017,** *114* (7), 1486.
27. Martirez, J. M. P.; Carter, E. A., Unraveling Oxygen Evolution on Iron-Doped β-Nickel Oxyhydroxide: The Key Role of Highly Active Molecular-like Sites. *Journal of the American Chemical Society* **2019,** *141* (1), 693-705.
28. Martirez, J. M. P.; Carter, E. A., Secondary Transition-Metal Dopants for Enhanced Electrochemical O2 Formation and Desorption on Fe-Doped β-NiOOH. *ACS Energy Letters* **2020,** *5* (3), 962-967.
29. Xiao, H.; Shin, H.; Goddard, W. A., Synergy between Fe and Ni in the optimal performance of (Ni,Fe)OOH catalysts for the oxygen evolution reaction. *Proceedings of the National Academy of Sciences* **2018,** *115* (23), 5872.
30. Shin, H.; Xiao, H.; Goddard, W. A., In Silico Discovery of New Dopants for Fe-Doped Ni Oxyhydroxide (Ni1–xFexOOH) Catalysts for Oxygen Evolution Reaction. *Journal of the American Chemical Society* **2018,** *140* (22), 6745-6748.
31. Tao, H. B.; Xu, Y.; Huang, X.; Chen, J.; Pei, L.; Zhang, J.; Chen, J. G.; Liu, B., A General Method to Probe Oxygen Evolution Intermediates at Operating Conditions. *Joule* **2019,** *3* (6), 1498-1509.
32. Yang, C.; Fontaine, O.; Tarascon, J. M.; Grimaud, A., Chemical Recognition of Active Oxygen Species on the Surface of Oxygen Evolution Reaction Electrocatalysts. *Angew Chem Int Ed Engl* **2017,** *56* (30), 8652-8656.
33. Lee, Y. M.; Kim, S.; Ohkubo, K.; Kim, K. H.; Nam, W.; Fukuzumi, S., Unified Mechanism of Oxygen Atom Transfer and Hydrogen Atom Transfer Reactions with a Triflic Acid-Bound Nonheme Manganese(IV)-Oxo Complex via Outer-Sphere Electron Transfer. *J Am Chem Soc* **2019,** *141* (6), 2614-2622.
34. Li, J.; Liao, H. J.; Tang, Y.; Huang, J. L.; Cha, L.; Lin, T. S.; Lee, J. L.; Kurnikov, I. V.; Kurnikova, M. G.; Chang, W. C.; Chan, N. L.; Guo, Y., Epoxidation Catalyzed by the Nonheme Iron(II)- and 2-Oxoglutarate-Dependent Oxygenase, AsqJ: Mechanistic Elucidation of Oxygen Atom Transfer by a Ferryl Intermediate. *J Am Chem Soc* **2020,** *142* (13), 6268-6284.
35. Holm; R., H., Metal-centered oxygen atom transfer reactions. *Chemical Reviews* **1987,** *87* (6), 1401-1449.
36. Gunay, A.; Theopold, K. H., C−H Bond Activations by Metal Oxo Compounds. *Chemical Reviews* **2010,** *110* (2), 1060-1081.

37. Zhou, M.; Crabtree, R. H., C–H oxidation by platinum group metal oxo or peroxo species. *Chem. Soc. Rev.* **2011,** *40* (4), 1875.
38. Lim, M. H.; Rohde, J.-U.; Stubna, A.; Bukowski, M. R.; Costas, M.; Ho, R. Y. N.; Münck, E.; Nam, W.; Que, L., An Fe-IV = O complex of a tetradentate tripodal nonheme ligand. *Proceedings of the National Academy of Sciences* **2003,** *100* (7), 3665.
39. Wang, H.; Casalongue, H. S.; Liang, Y.; Dai, H., Ni(OH)2 Nanoplates Grown on Graphene as Advanced Electrochemical Pseudocapacitor Materials. *Journal of the American Chemical Society* **2010,** *132* (21), 7472-7477.
40. Trotochaud, L.; Young, S. L.; Ranney, J. K.; Boettcher, S. W., Nickel-iron oxyhydroxide oxygen-evolution electrocatalysts: the role of intentional and incidental iron incorporation. *Journal of the American Chemical Society* **2014,** *136* (18), 6744-53.
41. Burstein, G. T., Special issue - A Century of Tafel's Equation: A Commemorative Issue of Corrosion Science. *Corrosion ence* **2005,** *47* (12), 2855-2856.
42. Bockris; O''M., J., Kinetics of Activation Controlled Consecutive Electrochemical Reactions: Anodic Evolution of Oxygen. *The Journal of Chemical Physics* **1956,** *24* (4), 817-827.
43. Bediako, D. K.; Surendranath, Y.; Nocera, D. G., Mechanistic studies of the oxygen evolution reaction mediated by a nickel-borate thin film electrocatalyst. *Journal of the American Chemical Society* **2013,** *135* (9), 3662-3674.
44. Yeo, B. S.; Klaus, S. L.; Ross, P. N.; Mathies, R. A.; Bell, A. T., Identification of hydroperoxy species as reaction intermediates in the electrochemical evolution of oxygen on gold. *Chemphyschem* **2010,** *11* (9), 1854-7.
45. Li, L.-F.; Li, Y.-F.; Liu, Z.-P., Oxygen Evolution Activity on NiOOH Catalysts: Four-Coordinated Ni Cation as the Active Site and the Hydroperoxide Mechanism. *ACS Catalysis* **2020,** *10* (4), 2581-2590.

# Supplementary Information

## Recognition of Surface Oxygen Intermediates on NiFe Oxyhydroxide Oxygen-evolving Catalysts by Homogeneous Oxidation Reactivity

Yaming Hao[1], Yefei Li[1*], Jianxiang Wu[1], Jinling Wang[2], Chenglin Jia[1], Tao Liu[1], Xuejing Yang[2], Zhipan Liu[1], Ming Gong[1*]

*Contains*

*28 figures*

*39 pages*

## Experimental section

### 1.1 Chemicals

All the reagents were used as received without further purification. Nickel nitrate hexahydrate ($Ni(NO_3)_2{\cdot}6H_2O$, 98%), iron nitrate nonahydrate ($Fe(NO_3)_3{\cdot}9H_2O$, AR), Cobalt nitrate hexahydrate ($Co(NO_3)_2{\cdot}6H_2O$, 99%), 4-diphenylphosphino-benzoic acid (TPP-COOH, > 97%), 4-methylbenzoic acid (Toluene-COOH, 98%), 4-hydroxybenzoic acid (Phenol-COOH, 99%), 4-vinyl-benzoic acid (Styrene-COOH, 97%), 4-(methylthio) benzoic acid (Thioanisole-COOH, > 97%), Deuterium oxide (98% atom % D, $D_2O$) and ethanol (≥ 99.7%) were purchased from Aladdin chemical reagent. Benzoic acid (Benzene-COOH, ≥ 99.5%) and potassium hydroxide (≥ 85 wt %, KOH) were obtained from Sinopharm Chemical Reagent Co. Ltd (China). Potassium deuteroxide solution (40 wt % in $D_2O$, 98% atom % D, KOD) were obtained from Sigma-Aldrich. Fluorene-4-carboxylic acid (Fluorene-COOH, 98%) were purchased from TCI (Shanghai). Anion exchange membrane were obtained from Fumatech (FAB-PK-130, Germany). Carbon fiber paper was purchased Hesen Electric Co. Ltd (HCP020N, China). Ultra-pure deionized water (18.2 $M\Omega{\cdot}cm^{-1}$, 25°C) was obtained from ELGA purification system.

### 1.2 Catalyst electrode preparation

The NiFe layered double hydroxide (LDH) catalyst precursors with different stoichiometric ratios of Ni/Fe were prepared by an electrodeposition method. Specifically, the electrodeposition was conducted in an electrochemical cell with a two-electrode configuration. 0.1 M $Ni(NO_3)_2{\cdot}6H_2O$ and x mM (x=0, 1, 5, 10, 20, 50) $Fe(NO_3)_3{\cdot}9H_2O$ aqueous solution was used as the electrolyte, and carbon fiber paper (CFP) with an active area of 1 cm × 1 cm was used as both the working electrode and counter electrode. The electrodeposition was carried out under a constant cathodic current density of 5 $mA{\cdot}cm^{-2}$ for 60 s (except for a current density of 20 $mA{\bullet}cm^{-2}$ for NiFe 50 due to the highly acidic electrolyte at high Fe contents). After electrodeposition, the electrode was washed with ethanol and ultra-pure deionized water, and further dried in a vacuum oven at 60 °C before its use for electrochemical measurement. The CoFe 5 LDH catalyst electrode was prepared under the identical process with the $Ni(NO_3)_2{\bullet}6H_2O$ precursor replaced by $Co(NO_3)_2{\bullet}6H_2O$.

**1.3 Electrochemical kinetics studies**

The electrochemical kinetics studies were performed on CHI 660E potentiostat (Shanghai Chenhua Instruments Co.) using a standard 100 mL H-type electrochemical cell with a three-electrode configuration. Graphite rod was used as the counter electrode; Ag/AgCl electrode was used as the reference electrode; the as-prepared NiFe LDH catalyst electrode was used as the working electrode. The electrolyte was mostly 1 M KOH aqueous solution unless specified. Cyclic voltammetry (CV) curves and linear sweep voltammetry (LSV) curves under the scan rates of 5 mV/s and 0.5 mV/s respectively were measured for the kinetic study. The CV and LSV curves were not iR-compensated for fair comparison of the conditions with and without probe, and most of the detailed resistance values are listed in Table S2 with others listed in Figure legends. For obtaining accurate Tafel slopes, all Tafel plots were iR-corrected. In a titration experiment, 5 mM reactive probes were added successively with each portion of 1 mM, and following each addition, we carried out a CV and a LSV measurement. All experiments were conducted at thermostatic water bath under a constant temperature of 30 °C. All measured potentials was converted to the reversible hydrogen electrode (RHE) according to the equation of E (RHE)=E (Ag/AgCl) +0.197+0.0591×pH. The H/D exchange experiment was carried out under almost identical conditions except that a 30 mL H-type electrochemical cell was used to minimize the use of KOD and $D_2O$. 1 M KOD in $D_2O$ was prepared by diluting 40 wt % KOD in $D_2O$ with $D_2O$, and the electrolyte with TPP-COOH addition was prepared before the measurement. The same electrode was used for both KOH and KOD experiments for reliable measurement, and the pH of the solution was calibrated by pH meter (PHSJ-3F, INESA Scientific Instrument Co. Ltd, China).

## 1.4 Characterization

### 1.4.1 FESEM

The electrodeposited NiFe LDH catalyst on carbon fiber paper (NiFe 5) was used as the sample. The field emission scanning electron microscopy（FESEM）images were recorded on the ZEISS MERLIN Compact.

### 1.4.2 *Operando* Raman spectroscopy

The Raman spectra were recorded by Raman spectrometer (Horiba Jobin Yvon) equipped with a 532 nm laser. The resolution is about 1.3 $cm^{-1}$ and 2400/mm grating was used. The acquisition time was set as 50 s and the spectral Raman shift range was set from 200-1800 $cm^{-1}$. The *operando* Raman spectroscopy was conducted by a custom-built three-electrode cell (EC-RAIR, Beijing Science Star technology Co. Ltd), with the Pt wire as the counter electrode, the Ag/AgCl electrode as the reference electrode and a roughened gold surface as the working electrode.

The NiFe 5 catalyst for the *operando* Raman spectroscopic study was prepared by a modified electrodeposition method on the roughened gold electrode. The electrodeposition condition was almost identical except that a lower current of 1 mA and a longer deposition duration of 120 s were used. The working electrode was further dried in vacuum at 30 °C after washing with ultra-pure deionized water and ethanol. Before the spectroscopic study, the electrodes were pre-activated by CV scans in 1 M KOH electrolyte to stabilize the electrode at a scan rate of 40 mV/s. The Raman spectra were collected afterwards by sweeping the electrode to the designated potential range. Before and after electrolysis of TPP-COOH dispersed in the 1M KOH was dropped on a glass-plate using the capillary with an inner diameter of 1mm.

### 1.4.3 Product analysis

The product analysis of the probe transformation during the electrochemical measurement was performed after the constant potential electrolysis of the electrolyte with 5 mM probes under the NiFe 5 catalyst and at 1.62 V vs RHE for 2 h. The electrolyte after the electrolysis was neutralized by dilute sulfuric acid to form white precipitate. The precipitate was further filtered and washed with ultra-pure deionized water to remove the residual acid. Finally, the precipitate was dried in the vacuum oven at 60 ℃, and re-dispersed in methanol and $CDCl_3$ for characterization purposes. As for control, the MS and NMR spectra of pure TPP-COOH were collected by dispersing in methanol and $CDCl_3$.

The product analysis was first conducted by the nano ESI-MS on the A LTQ XL Orbitrap hybrid instrument (Thermo Fisher Scientific). The product dissolved in methanol was directly fed to the instrument. The MS inlet temperature was 275 ℃ and spray voltage of negative 1.5 kV was used. The distance between the tip of the spray emitter and ion transfer capillary to the MS was held constant at ca. 1 mm.

The product dissolved in $CDCl_3$ was further characterized by NMR spectroscopy on Avance III HD 400 MHZ Bruker.

**1.5 Calculation Details**

All density functional theory (DFT) calculations were carried out within the periodic plane wave framework as implemented in Vienna ab initio simulation package (VASP).[1] The electron-ion interaction was represented by the projector augmented wave (PAW), and the kinetic energy cutoff of plane wave was set as 400 eV. The PBE functional with on-site Coulomb repulsion was utilized[2-3], and the effective U-J terms were 4.3 and 5.5 eV for Fe and Ni ions, as suggested in previous literature.[4-8] For all systems, the spin-polarization has been considered to identify the true ground state electronic configuration. For γ-NiOOH, the spin ordering in the ground state is ferromagnetic within the $NiO_2$ sheets.[6, 9] To model the γ-NiOOH ($\bar{1}2\bar{1}0$) surface, we used a slab model with 8 Ni layers and a (3×3) surface supercell with dimensions 14.02 Å×15.56 Å, as shown in Figure S22; altogether, this corresponds to 303 atoms per unit cell. Due to the large size of this cell, the k-point mesh was restricted to the Γ point. The geometry convergence criterion was set as 0.08 eV/Å for the maximal component of force. Transition states of the O-O coupling were determined using the Dimer method.[10-11]

The solvation effect due to the long-range electrostatic interaction was modeled by a periodic continuum solvation model with modified Poisson-Boltzmann equation.[12-14] The computational hydrogen electrode (CHE) approach is utilized to assess the thermodynamics of OER on γ-NiOOH surfaces, which estimates the electrochemical potential of $H^+$ and $e^-$ pair via the electrochemical equilibrium in Standard Hydrogen Electrode (SHE), which is

$$\mu[H^+] + \mu^{SHE}[e^-] = 1/2G^0[H_2] \qquad (1)$$

$\mu[H^+]$ is the electrochemical potential of a proton; $G^0[H_2]$ the standard Gibbs free energies of hydrogen; $\mu^{SHE}[e^-]$ the electrochemical potential of an electron in SHE. The detail description of CHE approach can be found in our previous papers.[15]

## 1.6 Mechanism Analysis

The rate-limiting step can often be correlated with the kinetics and reactivity parameters in the mechanistic study. Some typical OER mechanisms and their corresponding parameters are listed as follow:

**Electrochemical metal peroxide pathway**

| Rate-limiting Step | Parameters | Resting State | OAT Reactivity | KIE effect* | Comments (Limiting current?) |
|---|---|---|---|---|---|
| $M + OH^- \rightarrow MOH + e^-$ | $n_b$=0, $n_r$=1, ν=1 $b = \frac{2RT}{F}$ =120 mV/decade | M | No | S# | No limiting current |
| $MOH + OH^- \rightarrow MO + H_2O + e^-$ | $n_b$=1, $n_r$=1, ν=1 $b = \frac{2RT}{3F}$ =40 mV/decade | MOH | No | P | No limiting current |
| $MO + OH^- \rightarrow MOOH + e^-$ | $n_b$=2, $n_r$=1, ν=1 $b = \frac{2RT}{5F}$ =24 mV/decade | MO | Yes | S | No limiting current |
| $MOOH + OH^- \rightarrow M + O_2 + H_2O + e^-$ | $n_b$=3, $n_r$=1, ν=1 $b = \frac{2RT}{7F}$ =17.1 mV/decade | MOOH | Less Likely | P | No limiting current |

*Assuming these reactions are proton-coupled electron transfer reactions without involving the limiting proton transfer or electron transfer reactions (often known as PTET or ETPT reactions)

#S=Secondary KIE, P=Primary KIE

**Metal oxide pathway**

| Rate-limiting Step | Parameters | Resting State | OAT Reactivity | KIE effect* | Comments (Limiting current?) |
|---|---|---|---|---|---|
| $M + OH^- \rightarrow MOH + e^-$ | $n_b$=0, $n_r$=1, ν=4 $b = \frac{2RT}{F}$ =120 mV/decade | M | No | S | No limiting current |
| $2MOH \rightarrow MO + M + H_2O$ | $n_b$=4, $n_r$=0, ν=2 $b = \frac{RT}{2F}$ =30 mV/decade | MOH | No | P | Possible limiting current |
| $2MO \rightarrow 2M + O_2$ | $n_b$=4, $n_r$=0, ν=1 $b = \frac{RT}{4F}$ =15 mV/decade | MO | Yes | No | Possible limiting current |

**Electrochemical metal oxide pathway**

| Rate-limiting Step | Parameters | Resting State | OAT Reactivity | KIE effect* | Comments (Limiting current?) |
|---|---|---|---|---|---|
| $M + OH^- \rightarrow MOH + e^-$ | $n_b$=0, $n_r$=1, ν=4 $b = \frac{2RT}{F}$ =120 mV/decade | M | No | S | No limiting current |
| $MOH + OH^- \rightarrow MO + H_2O + e^-$ | $n_b$=2, $n_r$=1, ν=2 $b = \frac{2RT}{3F}$ =40 mV/decade | MOH | No | P | No limiting current |
| $2MO \rightarrow 2M + O_2$ | $n_b$=4, $n_r$=0, ν=1 $b = \frac{RT}{4F}$ =15 mV/decade | MO | Yes | No | Possible limiting current |

**Lattice oxygen pathway**

| Rate-limiting Step | Parameters | Resting State | OAT Reactivity | KIE effect* | Comments (Limiting current?) |
|---|---|---|---|---|---|
| $M + OH^- \rightarrow MOH + e^-$ | $n_b$=0, $n_r$=1, ν=4 $b = \frac{2RT}{F}$ =120 mV/decade | M | No | S | No limiting current |
| $MOH + OH^- \rightarrow MO + H_2O + e^-$ | $n_b$=1, $n_r$=1, ν=1 $b = \frac{2RT}{3F}$ =40 mV/decade | MOH | No | P | No limiting current |
| $MO + O_{lat} \rightarrow 2M + O_2$ | $n_b$=2, $n_r$=0, ν=1 $b = \frac{RT}{2F}$ =30 mV/decade | MO | Yes | S or No | Possible limiting current |

One can often derive the mechanism according to these kinetics and reactivity parameters; however, since these mechanisms are simplified models, some deviation from the real mechanism could be possible when using this system. Two important examples are:

1) The initial state may not be under-coordinated or water-coordinated surfaces or the M mentioned in previous tables, and sometimes M-OH on hydroxide surfaces can be directly used as the initial state. The same rate-determining step can generate different Tafel slopes. Therefore, mechanism deduction simply from Tafel slopes is not accurate and sufficient knowledge about the initial state is IMPORTANT!

| **M as the initial state** | | **M-OH as the initial state** | |
|---|---|---|---|
| **Rate-limiting Step** | **Parameters** | **Rate-limiting Step** | **Parameters** |
| $M + OH^- \rightarrow MOH + e^-$ | $n_b$=0, $n_r$=1, ν=1<br>$b = \frac{2RT}{F}$<br>=120 mV/decade | $MOH + OH^- \rightarrow MO + H_2O + e^-$ | $n_b$=0, $n_r$=1, ν=1<br>$b = \frac{2RT}{F}$<br>=120 mV/decade |
| $MOH + OH^- \rightarrow MO + H_2O + e^-$ | $n_b$=1, $n_r$=1, ν=1<br>$b = \frac{2RT}{3F}$<br>=40 mV/decade | $MO + OH^- \rightarrow MOOH + e^-$ | $n_b$=1, $n_r$=1, ν=1<br>$b = \frac{2RT}{3F}$<br>=40 mV/decade |
| $MO + OH^- \rightarrow MOOH + e^-$ | $n_b$=2, $n_r$=1, ν=1<br>$b = \frac{2RT}{5F}$<br>=24 mV/decade | $MOOH + OH^- \rightarrow M + O_2 + H_2O + e^-$ | $n_b$=2, $n_r$=1, ν=1<br>$b = \frac{2RT}{5F}$<br>=24 mV/decade |
| $MOOH + OH^- \rightarrow M + O_2 + H_2O + e^-$ | $n_b$=3, $n_r$=1, ν=1<br>$b = \frac{2RT}{7F}$<br>=17.1 mV/decade | | |

2) MOH, MO and etc. are simplied models, and different –OH and –O generated on different sites could have distinct reactivities. We summarize the OAT and HAT reactivities of the possible oxygen intermediates (e.g. atop sites and bridging sites) according to literature and our understanding to help us derive the OER mechanism.

| Oxygen Intermediates | Reactivity | |
|---|---|---|
| | OAT | HAT |
| 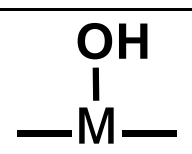 | × | $M\text{-}OH + RH \rightarrow M + H_2O + R\cdot$ |
| 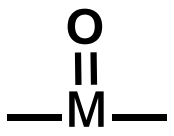 | M=O + S $\rightarrow$ M + S=O | M=O + RH $\rightarrow$ M-OH + R· |
| 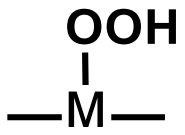 | × | M-OOH + RH $\rightarrow$ M=O + R· + $H_2O$<br>(Less likely) |
| 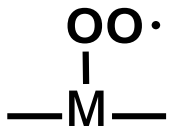 | M-OO· + S $\rightarrow$ M-O· + S=O | M-OO· + RH $\rightarrow$ M-OOH + R· |
| 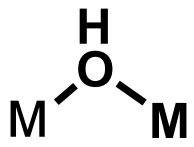 | × | M-OH-M + RH $\rightarrow$ M M + $H_2O$ + R·<br>(Less likely) |
| 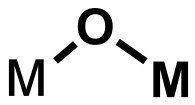 | M-O-M + S$\rightarrow$ M M + S=O<br>(Less likely) | M-O-M + RH $\rightarrow$ M-OH-M + + R· |
| 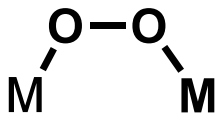 | × | × |
| 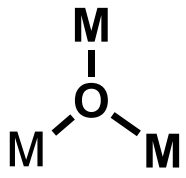 | × | M-O-$(M)_2$ + RH $\rightarrow$ M-OH-$(M)_2$ + R· |
| 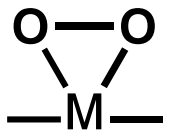 | M-$(O_2)$ + S$\rightarrow$ M=O + S=O | M-$(O_2)$ + RH $\rightarrow$ M-OOH + R· |

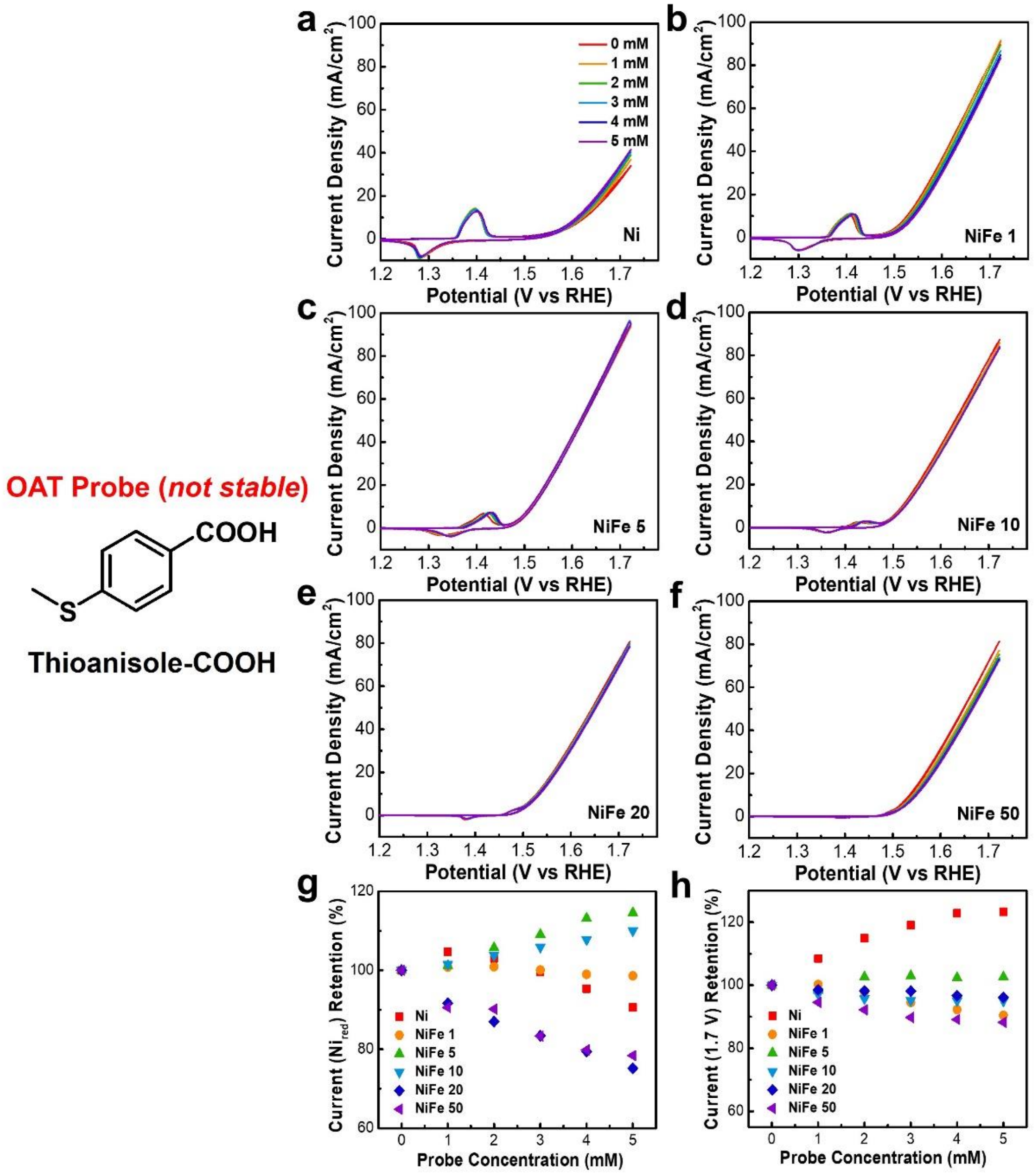

**Figure S1**. a-f) Cyclic voltammetry curves of NiFe LDH catalysts with different Fe/Ni ratios in 1 M KOH under the titration of 4-(methylthio)benzoic acid (Thioanisole-COOH). The scan rate is 5 mV/s. The insets show the corresponding Ni redox features. g) The trend of $Ni^{\delta+}$ reduction peak current over different Thioanisole-COOH concentrations. h) The trend of OER current at 1.7 V vs RHE over different Thioanisole-COOH concentrations.

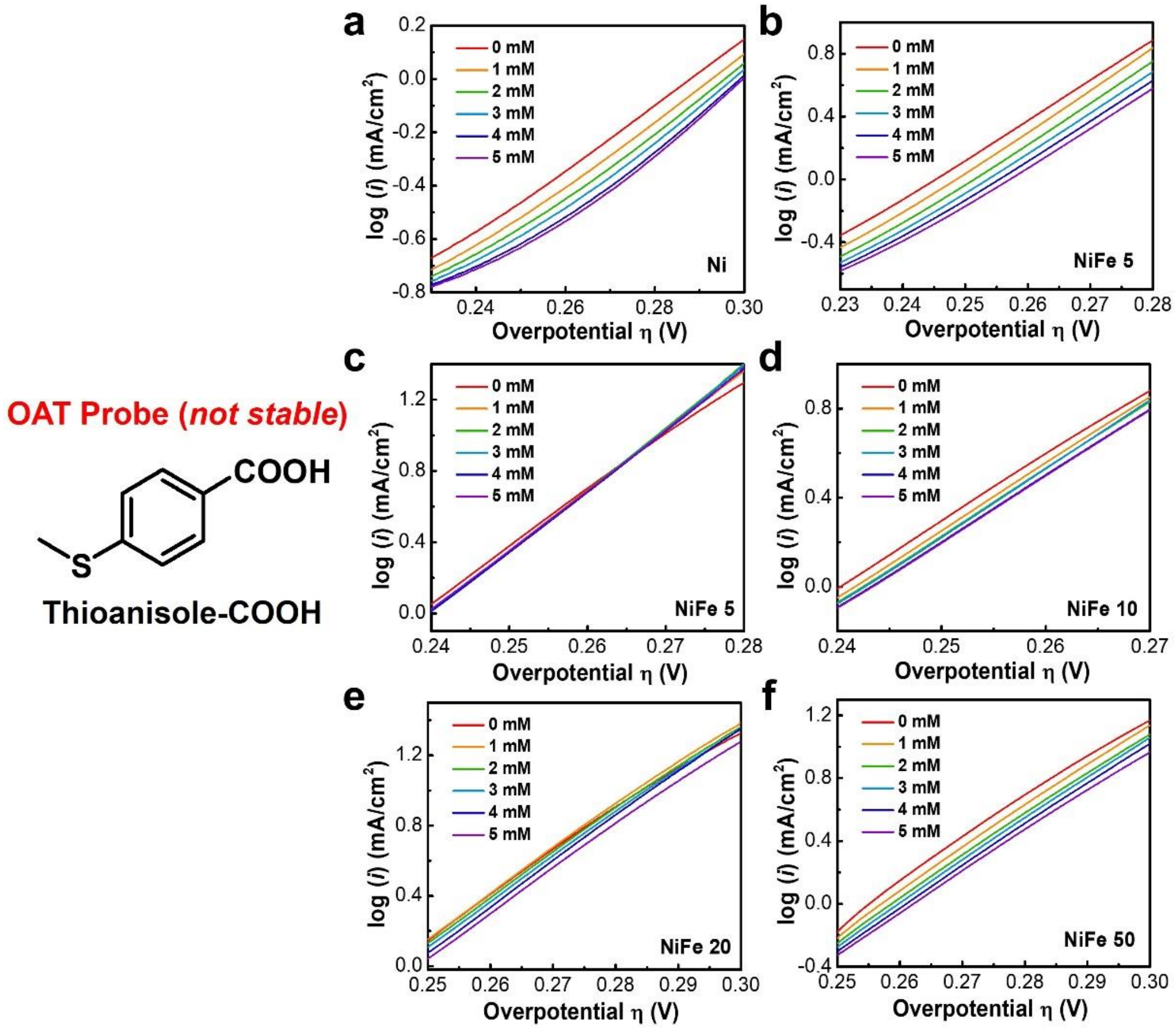

**Figure S2**. a-f) Tafel plots of NiFe LDH catalysts with different Fe/Ni ratios in 1 M KOH under the titration of 4-(methylthio)benzoic acid (Thioanisole-COOH). The scan rate is 0.5 mV/s.

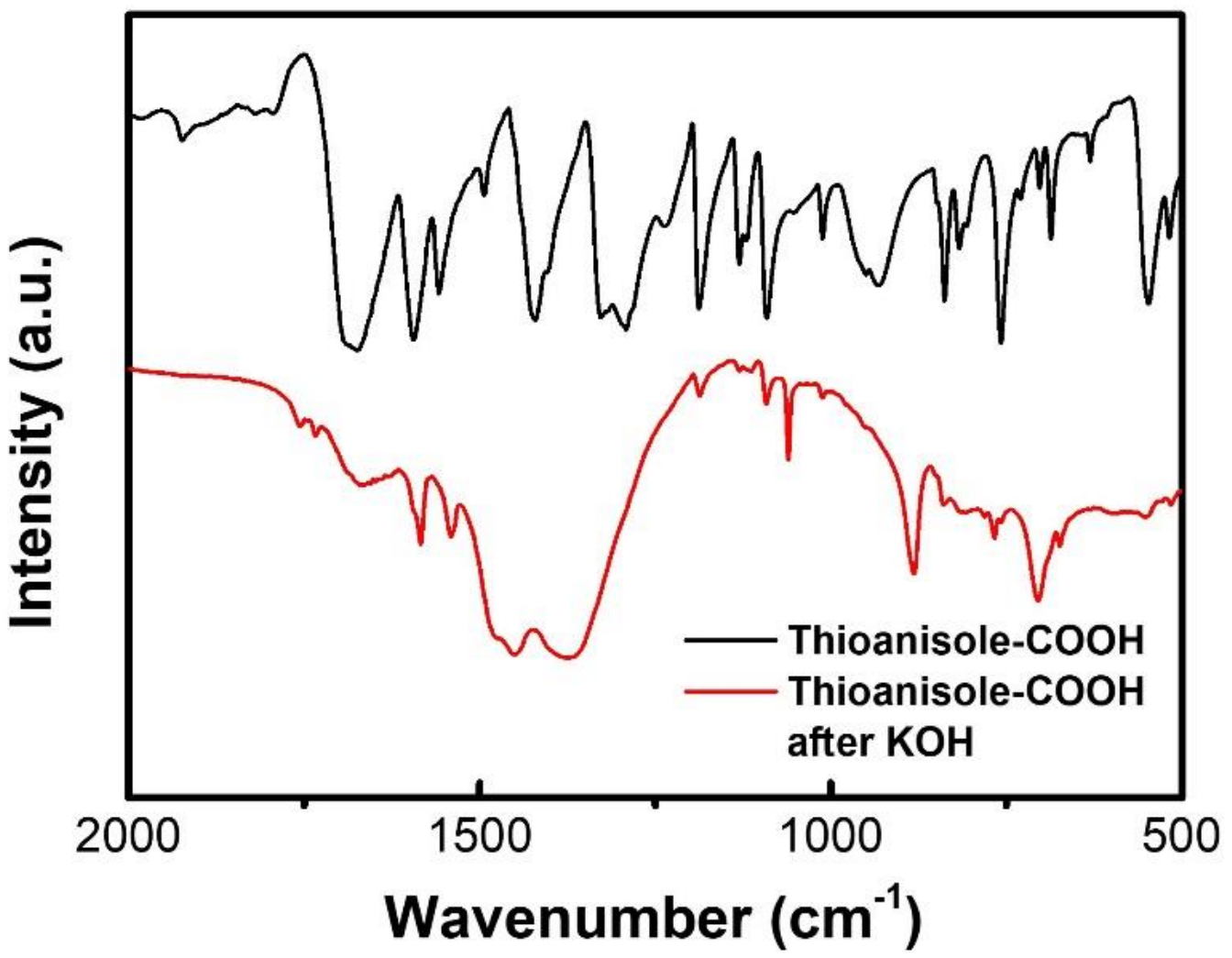

**Figure S3**. The FT-IR spectra of the 4-(methylthio) benzoic acid (Thioanisole-COOH) before and after mixed with KOH, demonstrating significant instability under alkaline conditions

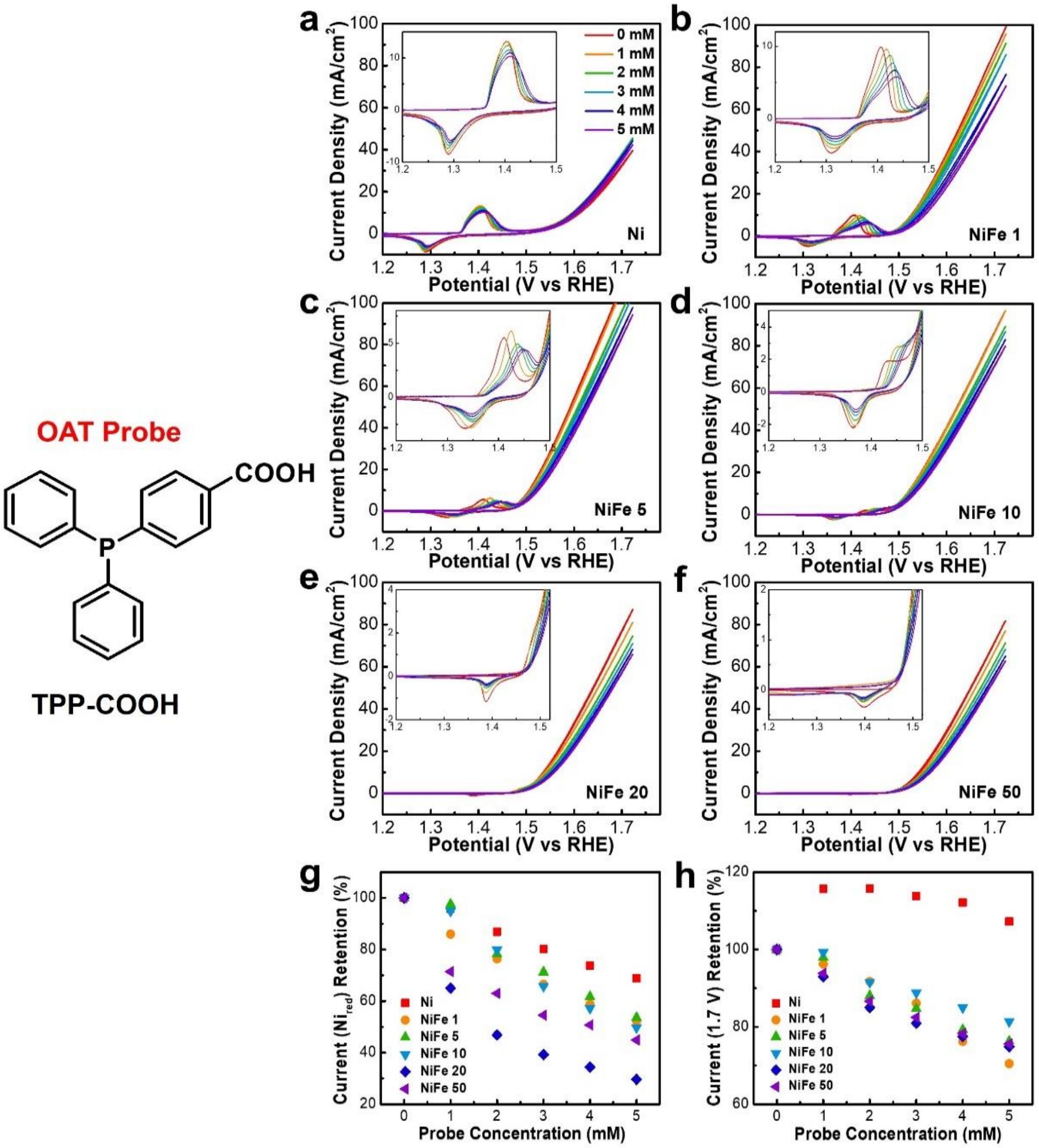

**Figure S4**. a-f) Cyclic voltammetry curves of NiFe LDH catalysts with different Fe/Ni ratios in 1 M KOH under the titration of 4-diphenylphosphino-benzoic acid (TPP-COOH). The scan rate is 5 mV/s. The insets show the corresponding Ni redox features. g) The trend of $Ni^{\delta+}$ reduction peak current over different TPP-COOH concentrations. h) The trend of OER current at 1.7 V vs RHE over different TPP-COOH concentrations. All NiFe-based catalysts showed drastic decrease of both $Ni^{\delta+}$ reduction current and OER current under increasing concentrations of TPP-COOH

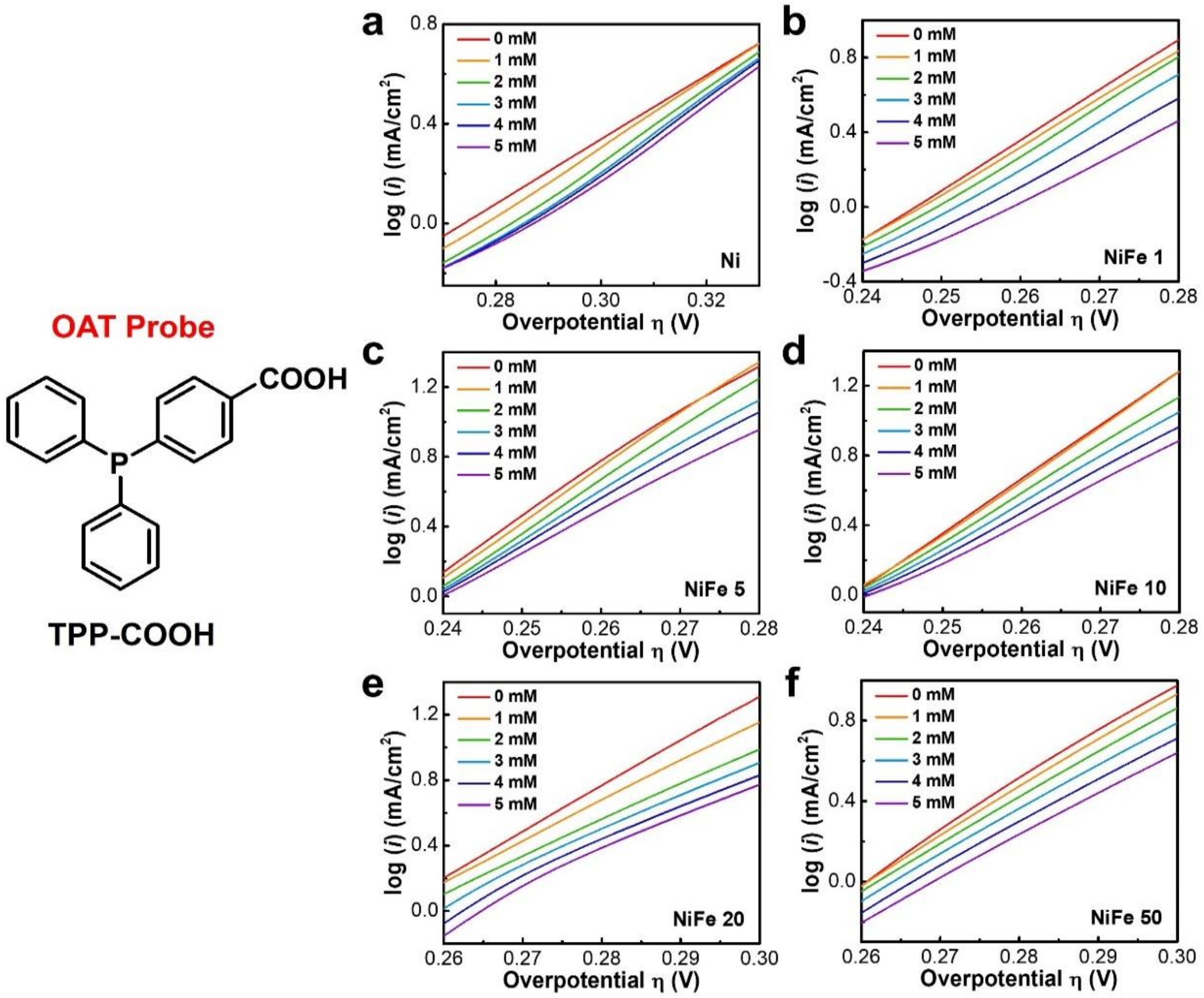

**Figure S5**. a-f) Tafel plots of NiFe LDH catalysts with different Fe/Ni ratios in 1 M KOH under the titration of 4-diphenylphosphino-benzoic acid (TPP-COOH). The scan rate is 0.5 mV/s. All NiFe-based catalysts showed significant increase of Tafel slopes upon the addition of TPP-COOH.

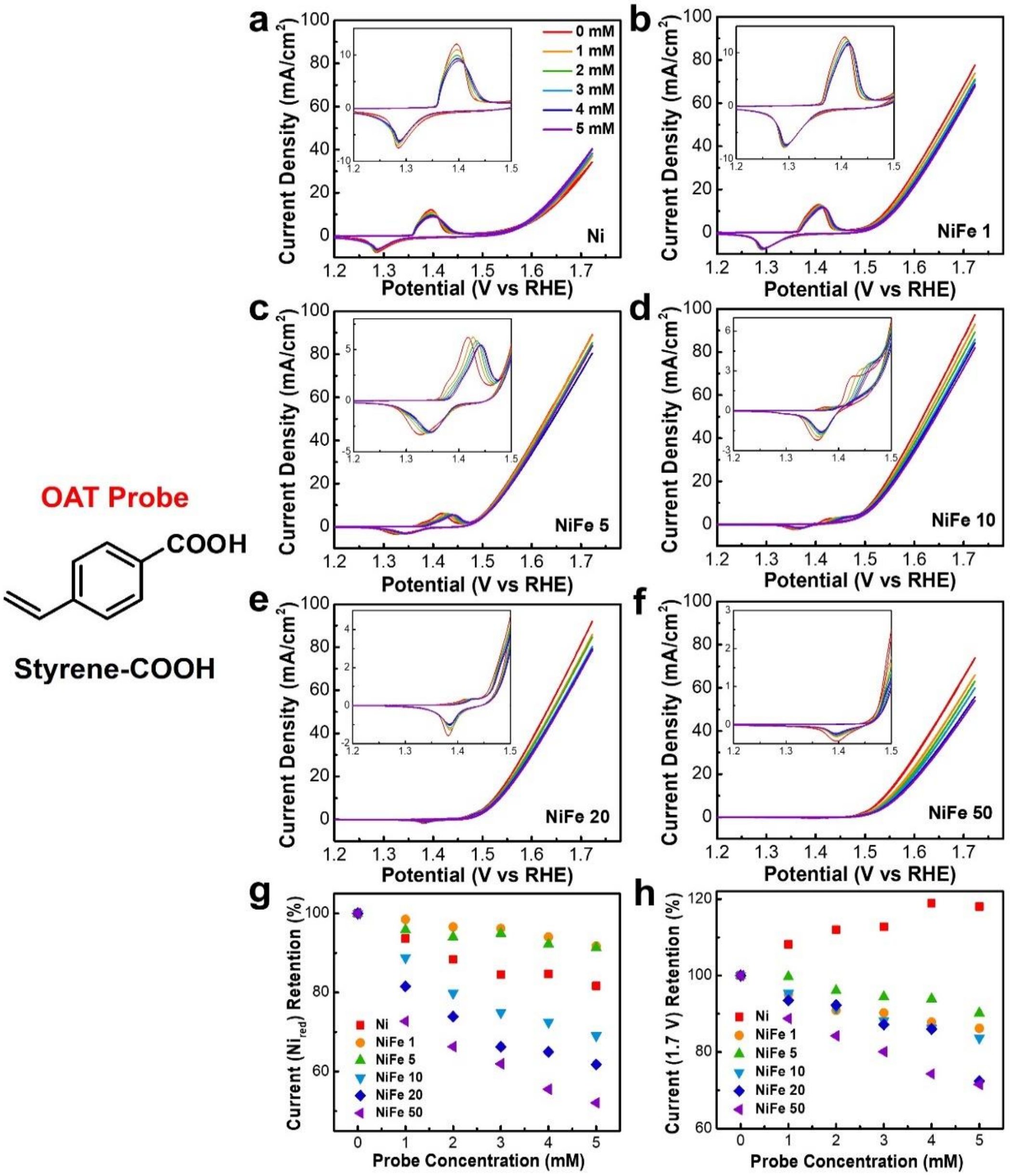

**Figure S6**. a-f) Cyclic voltammetry curves of NiFe LDH catalysts with different Fe/Ni ratios in 1 M KOH under the titration of 4-vinyl-benzoic acid (Styrene-COOH). The scan rate is 5 mV/s. The insets show the corresponding Ni redox features. g) The trend of $Ni^{\delta+}$ reduction peak current over different Styrene-COOH concentrations. h) The trend of OER current at 1.7 V vs RHE over different Styrene-COOH concentrations.

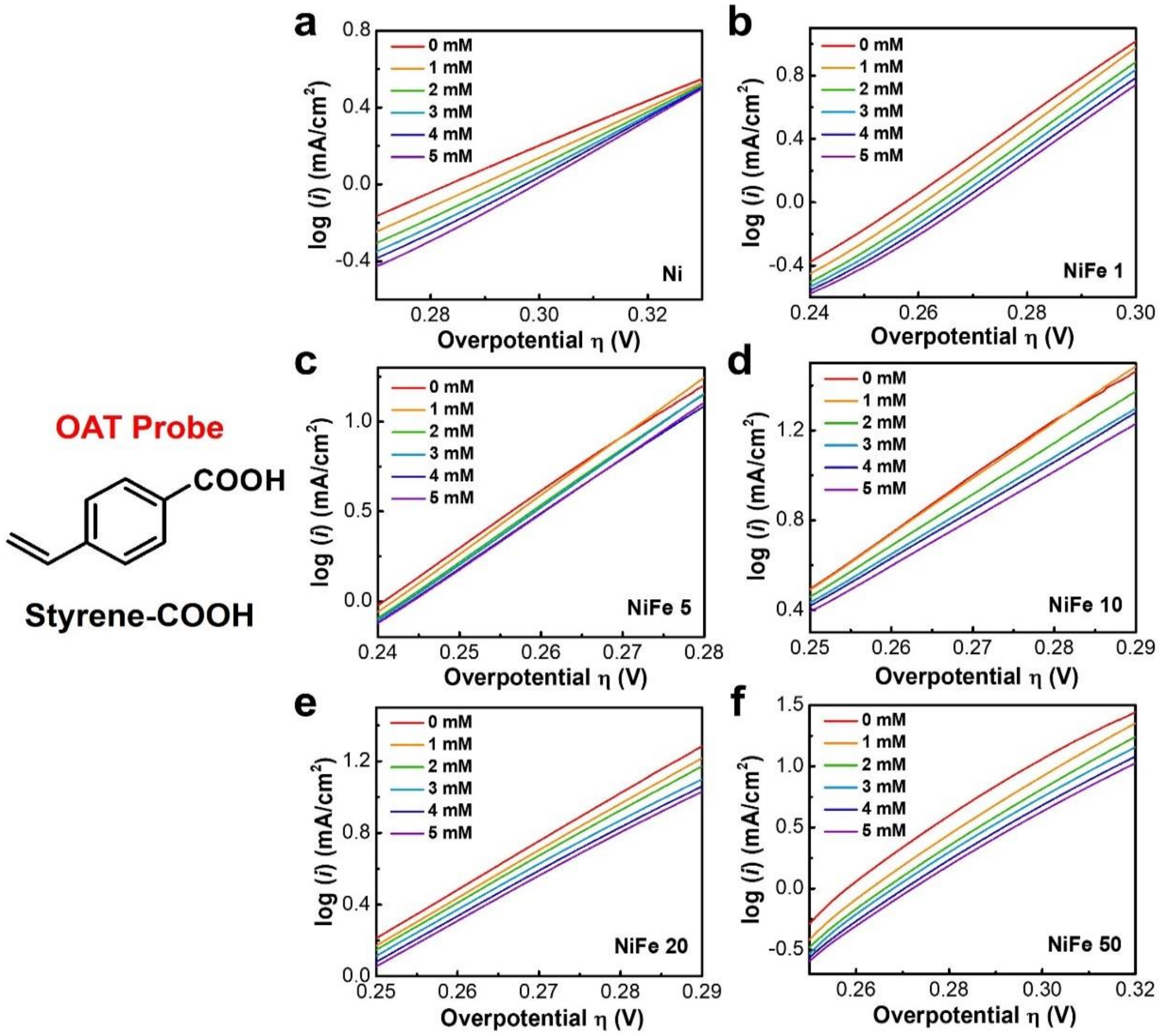

**Figure S7**. a-f) Tafel plots of NiFe LDH catalysts with different Fe/Ni ratios in 1 M KOH under the titration of 4-vinyl-benzoic acid (Styrene-COOH). The scan rate is 0.5 mV/s.

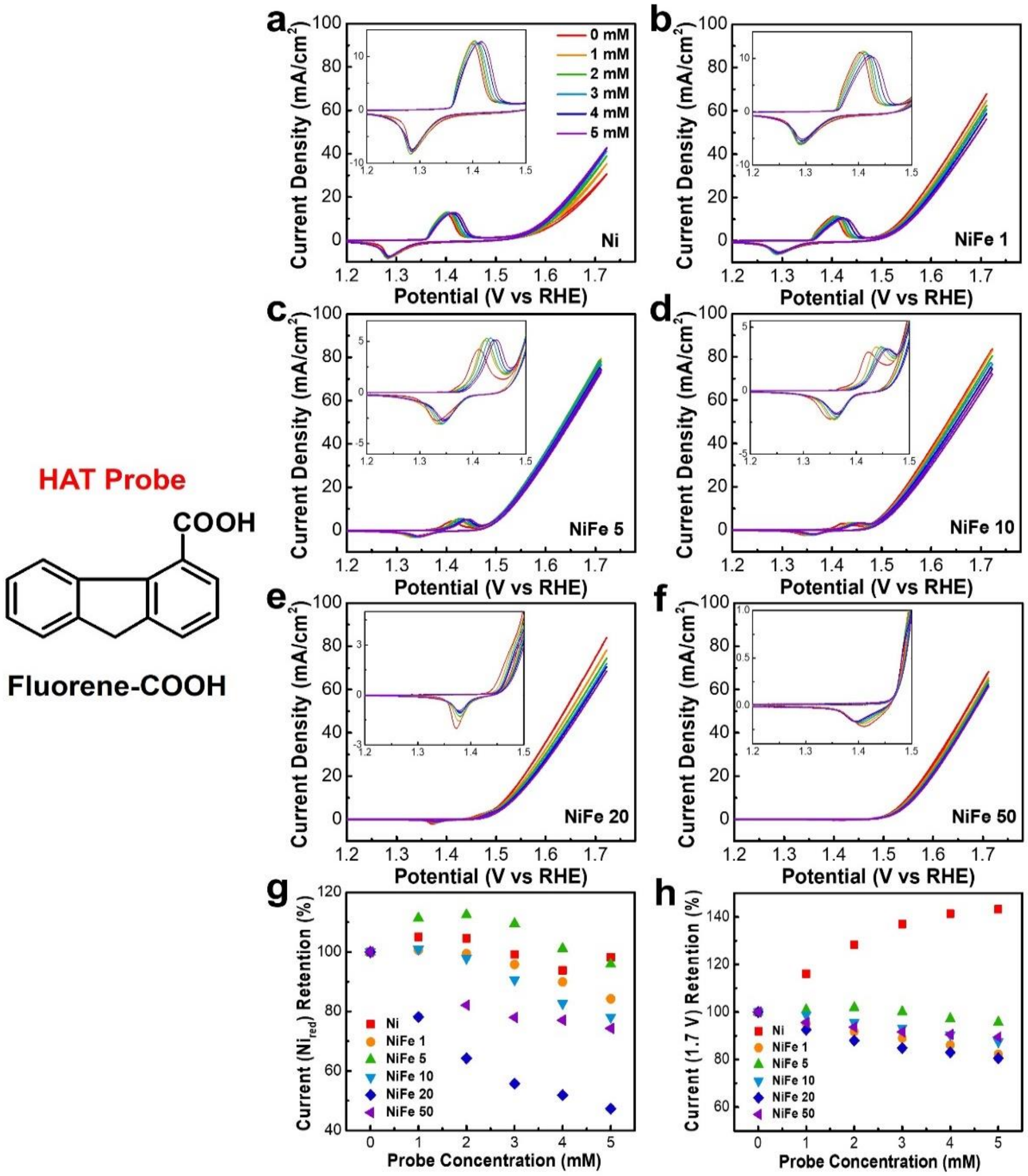

**Figure S8**.

a-f) Cyclic voltammetry curves of NiFe LDH catalysts with different Fe/Ni ratios in 1 M KOH under the titration of Fluorene-4-carboxylic acid (Fluorene-COOH). The scan rate is 5 mV/s. The insets show the corresponding Ni redox features. g) The trend of $Ni^{\delta+}$ reduction peak current over different Fluorene-COOH concentrations. h) The trend of OER current at 1.7 V vs RHE over different Fluorene-COOH concentrations.

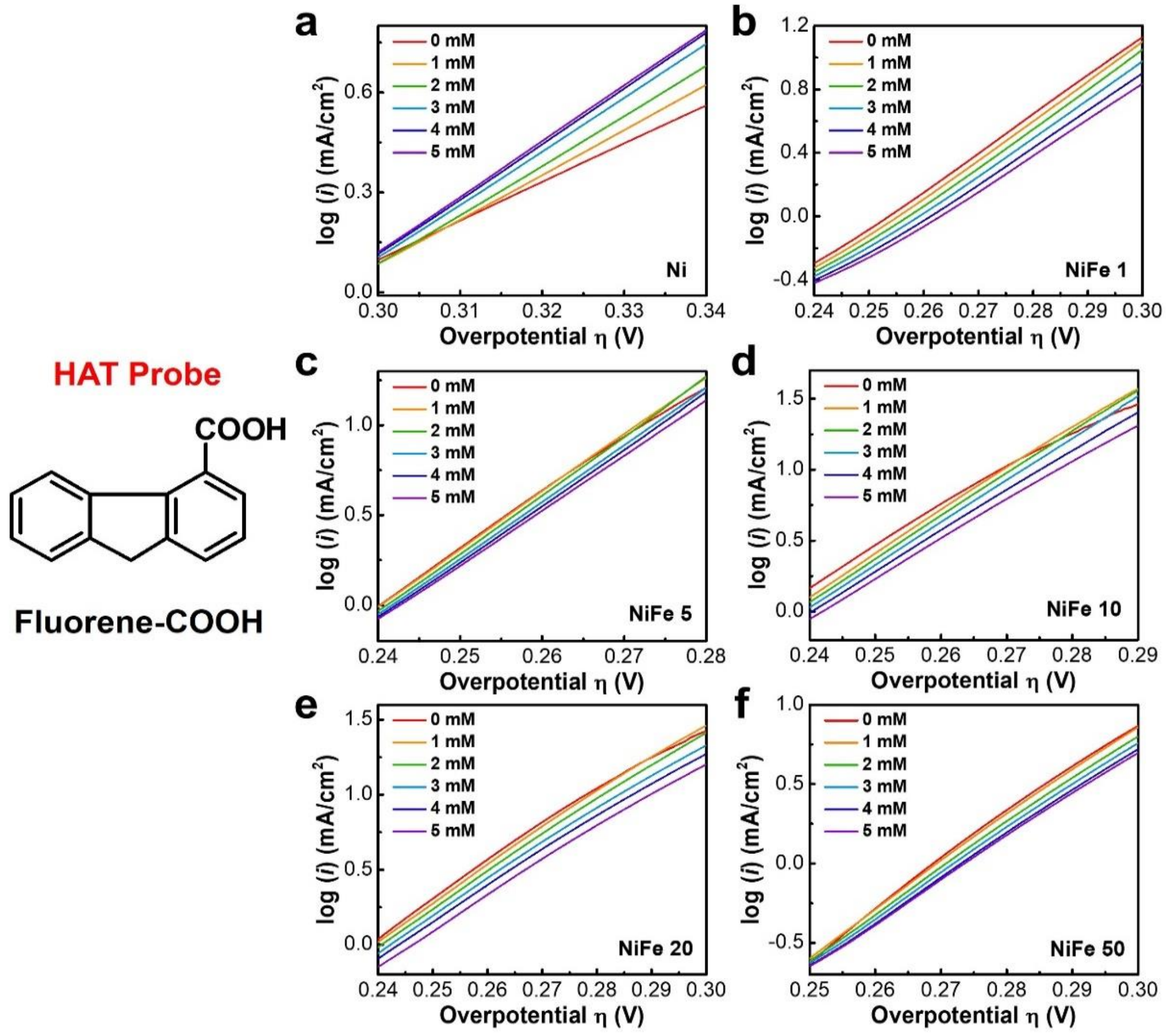

**Figure S9**. a-f) Tafel plots of NiFe LDH catalysts with different Fe/Ni ratios in 1 M KOH under the titration of Fluorene-4-carboxylic acid (Fluorene-COOH). The scan rate is 0.5 mV/s.

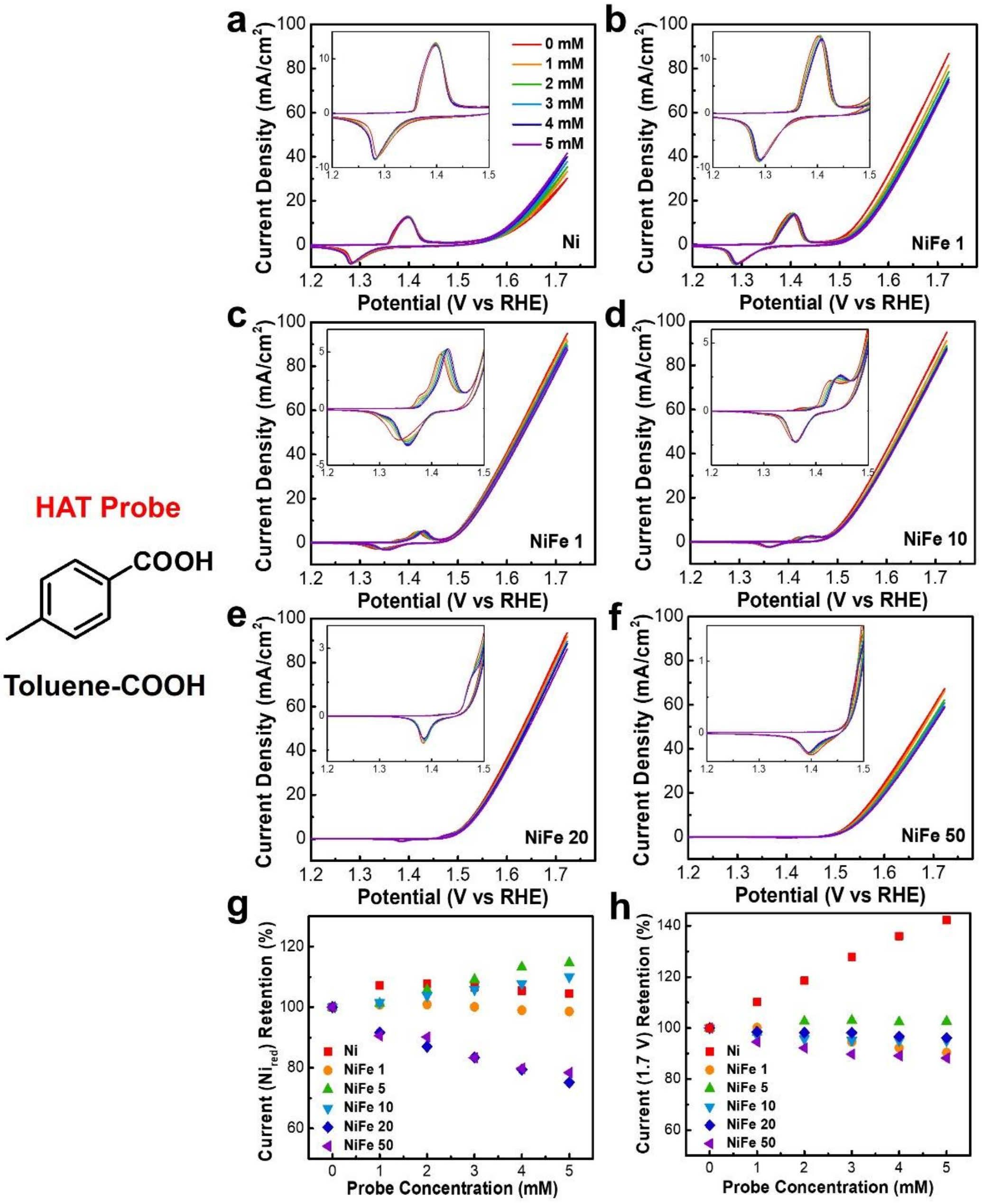

**Figure S10**. a-f) Cyclic voltammetry curves of NiFe LDH catalysts with different Fe/Ni ratios in 1 M KOH under the titration of 4-methylbenzoic acid (Toluene-COOH). The scan rate is 5 mV/s. The insets show the corresponding Ni redox features. g) The trend of $Ni^{\delta+}$ reduction peak current over different Toluene-COOH concentrations. h) The trend of OER current at 1.7 V vs RHE over different Toluene-COOH concentrations.

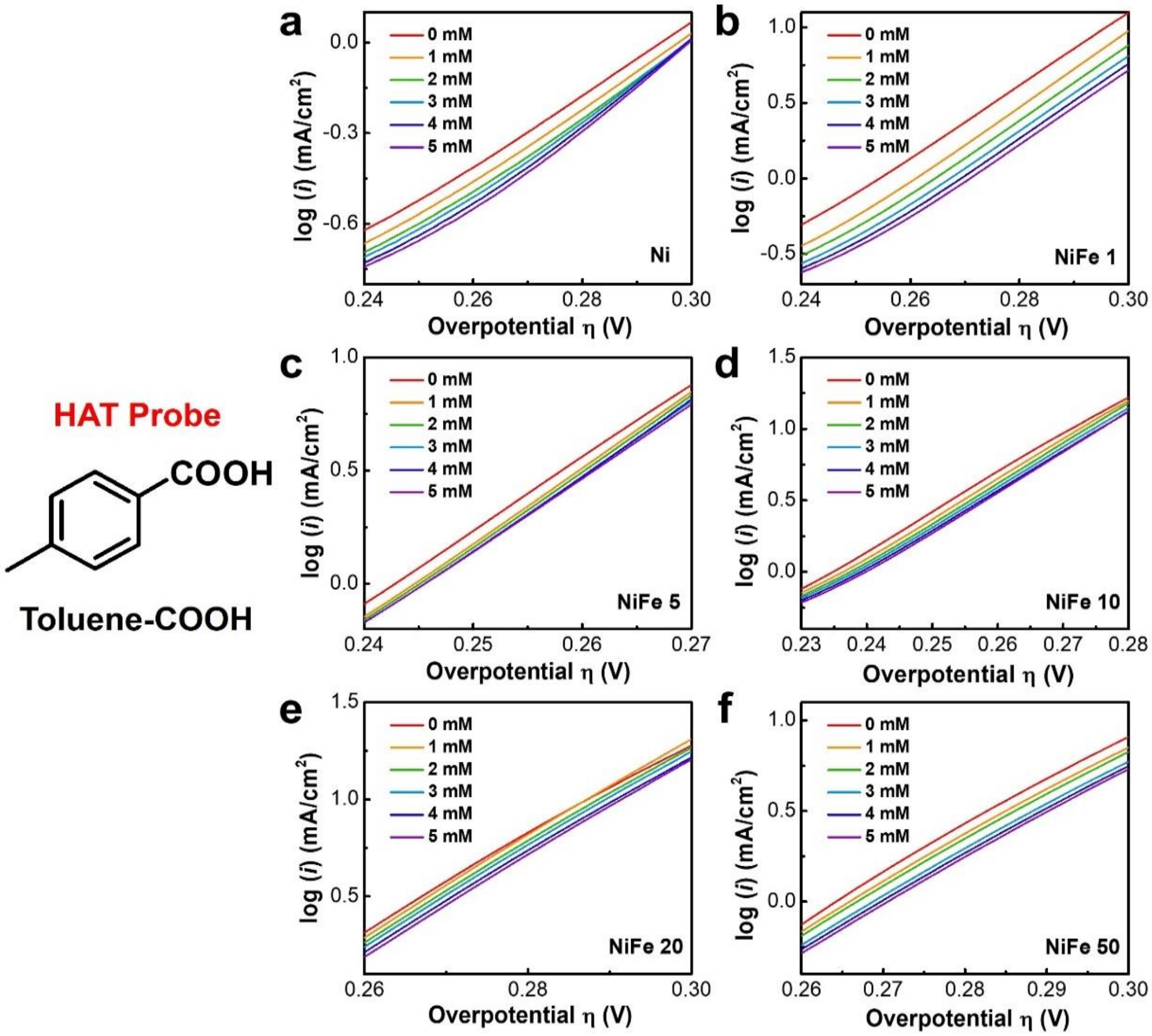

**Figure S11**. a-f) Tafel plots of NiFe LDH catalysts with different Fe/Ni ratios in 1 M KOH under the titration of 4-methylbenzoic acid (Toluene-COOH). The scan rate is 0.5 mV/s.

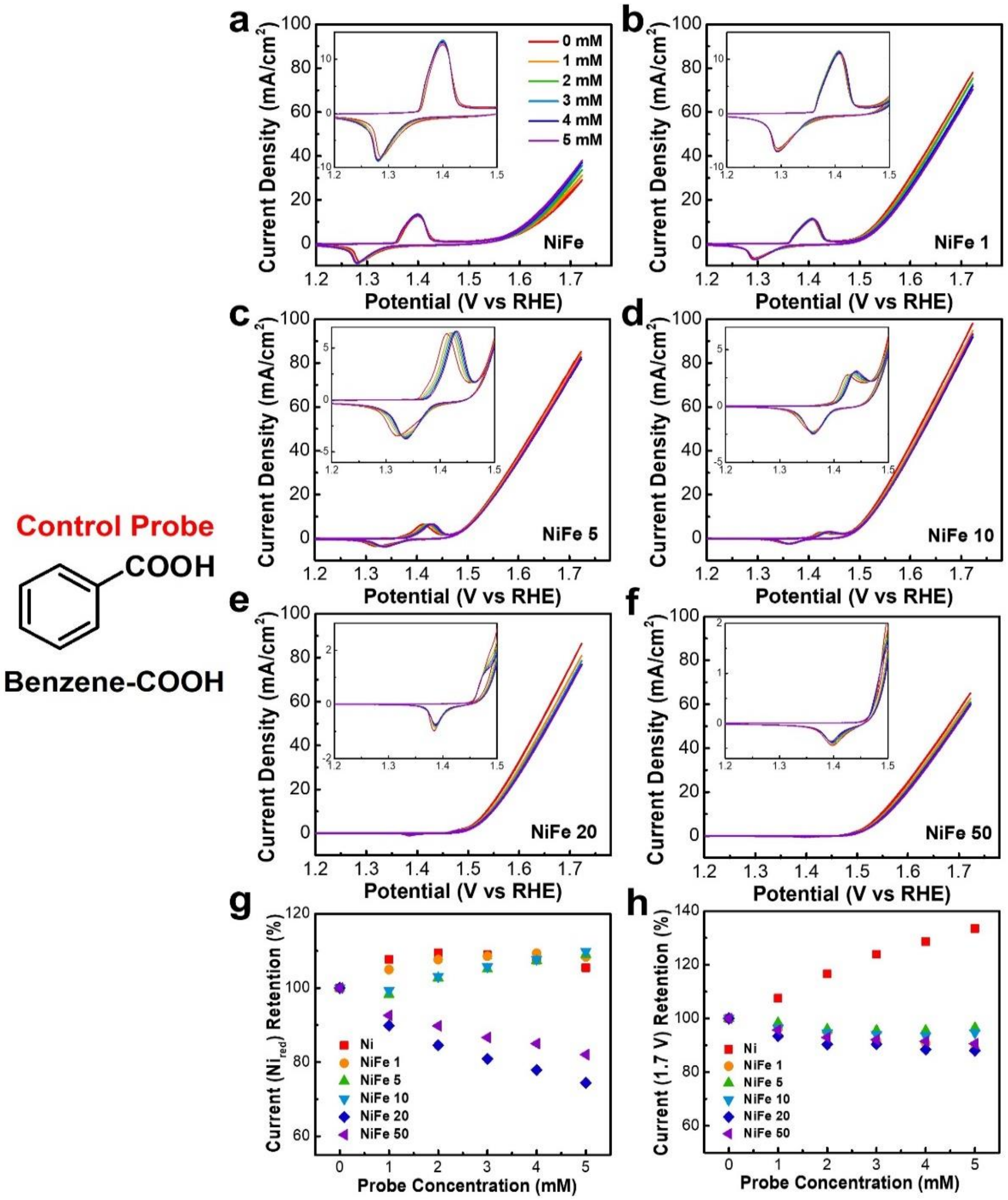

**Figure S12**. a-f) Cyclic voltammetry curves of NiFe LDH catalysts with different Fe/Ni ratios in 1 M KOH under the titration of benzoic acid (Benzene-COOH). The scan rate is 5 mV/s. The insets show the corresponding Ni redox features. g) The trend of $Ni^{\delta+}$ reduction peak current over different Benzene-COOH concentrations. h) The trend of OER current at 1.7 V vs RHE over different Benzene-COOH concentrations.

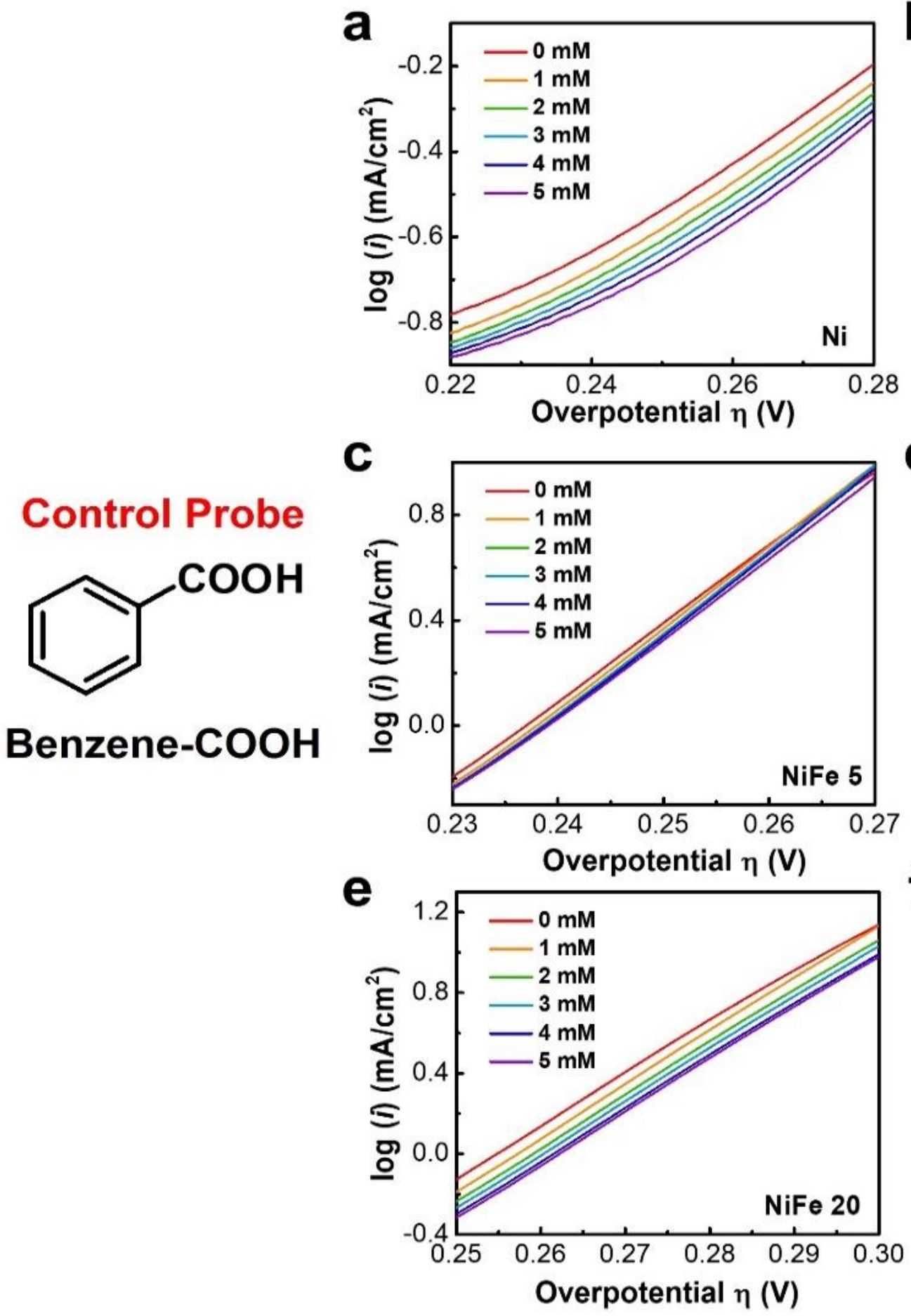

**Figure S13**. a-f) Tafel plots of NiFe LDH catalysts with different Fe/Ni ratios in 1 M KOH under the titration of benzoic acid (Benzene-COOH). The scan rate is 0.5 mV/s.

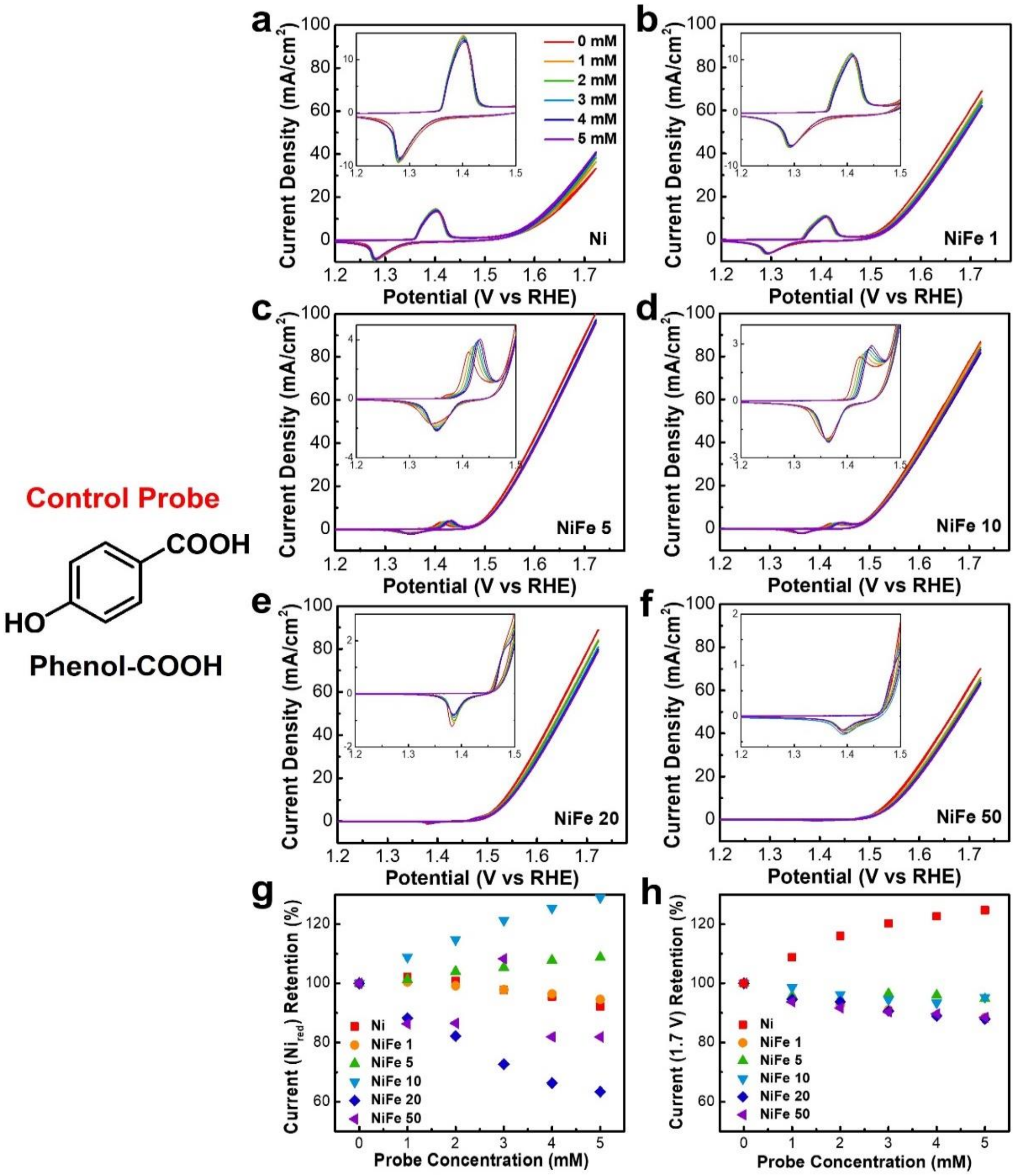

**Figure S14**. a-f) Cyclic voltammetry curves of NiFe LDH catalysts with different Fe/Ni ratios in 1 M KOH under the titration of 4-hydroxybenzoic acid (Phenol-COOH). The scan rate is 5 mV/s. The insets show the corresponding Ni redox features. g) The trend of $Ni^{\delta+}$ reduction peak current over different Phenol-COOH concentrations. h) The trend of OER current at 1.7 V vs RHE over different Phenol-COOH concentrations.

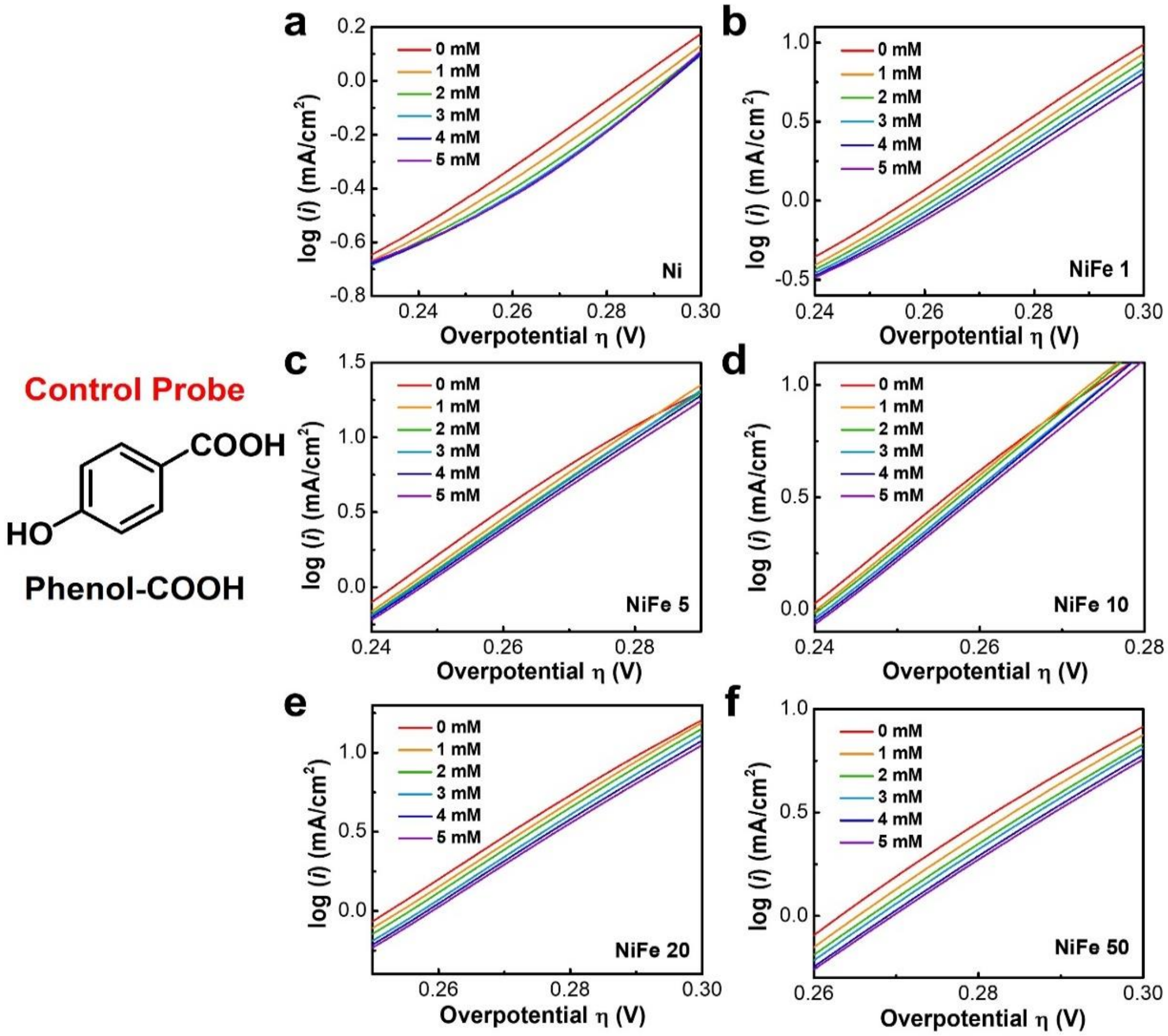

**Figure S15**. a-f) Tafel plots of NiFe LDH catalysts with different Fe/Ni ratios in 1 M KOH under the titration of 4-hydroxybenzoic acid (Phenol-COOH). The scan rate is 0.5 mV/s.

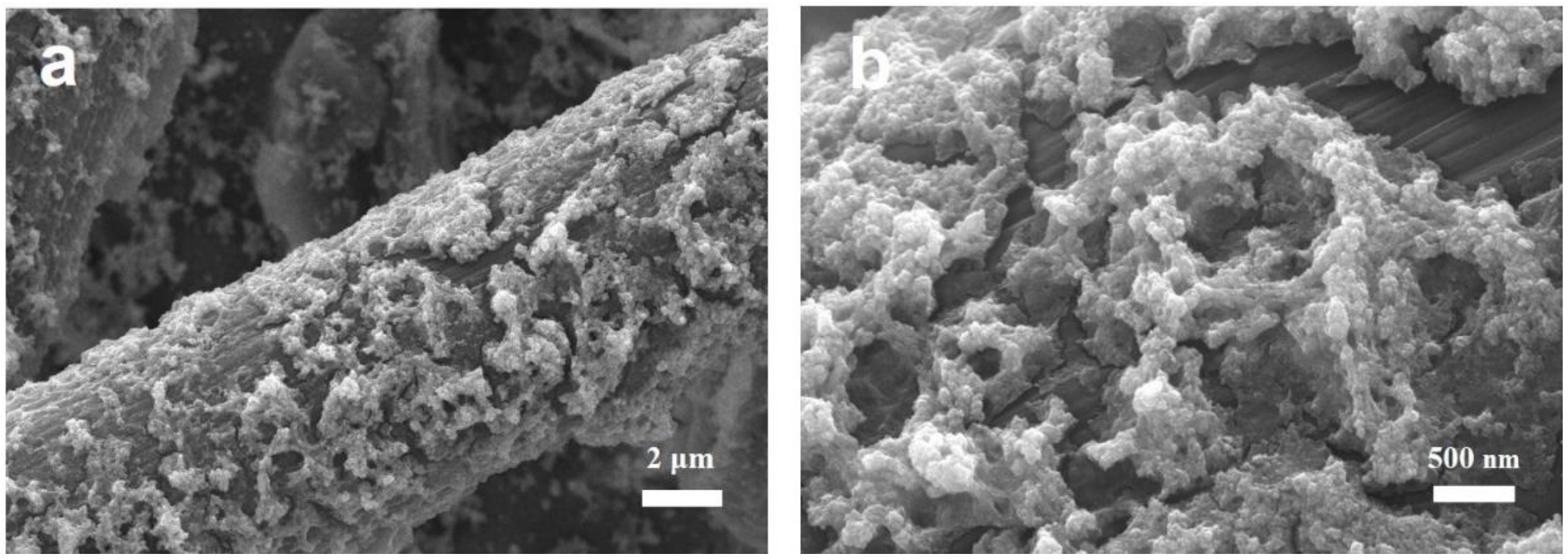

**Figure S16**. SEM images of NiFe 5 LDH catalyst electrodeposited on the carbon fiber paper electrode

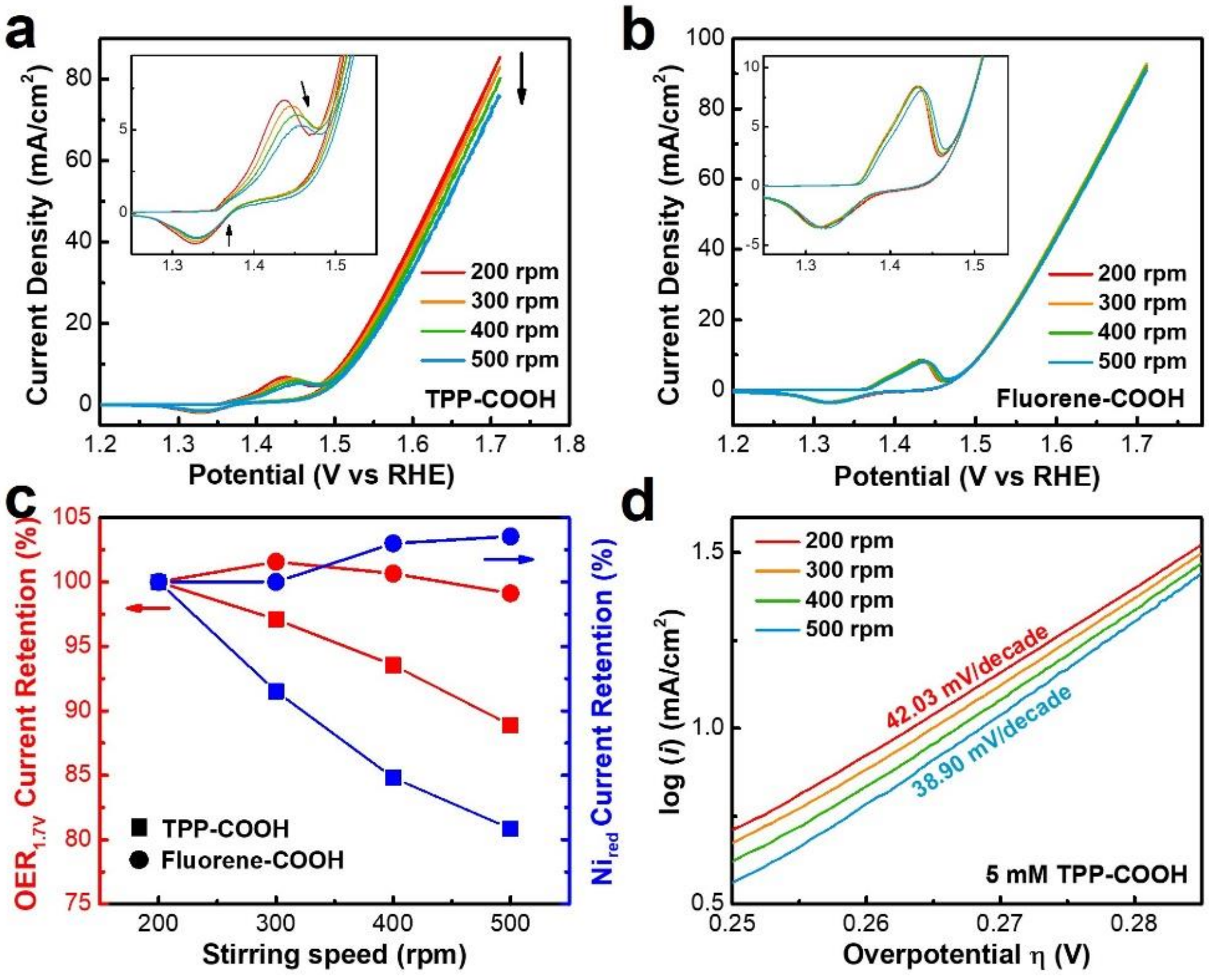

**Figure S17**. a-b) Cyclic voltammetry curves of the NiFe 5 (Fe/Ni=5%) catalysts in 1 M KOH with 5 mM probes, a) TPP-COOH (R=2.2Ω), b) Fluorene-COOH (R=2.2Ω), under different stirring conditions. The insets show the Ni redox features under different stirring conditions. c) The change in the two parameters of $Ni^{\delta+}$ reduction current and OER current at 1.7 V vs RHE over NiFe 5 catalyst under the addition of 5 mM probe and different stirring conditions. d) Tafel plot of the NiFe 5 (Fe/Ni=5%) catalysts in 1 M KOH with 5 mM TPP-COOH over different stirring conditions.

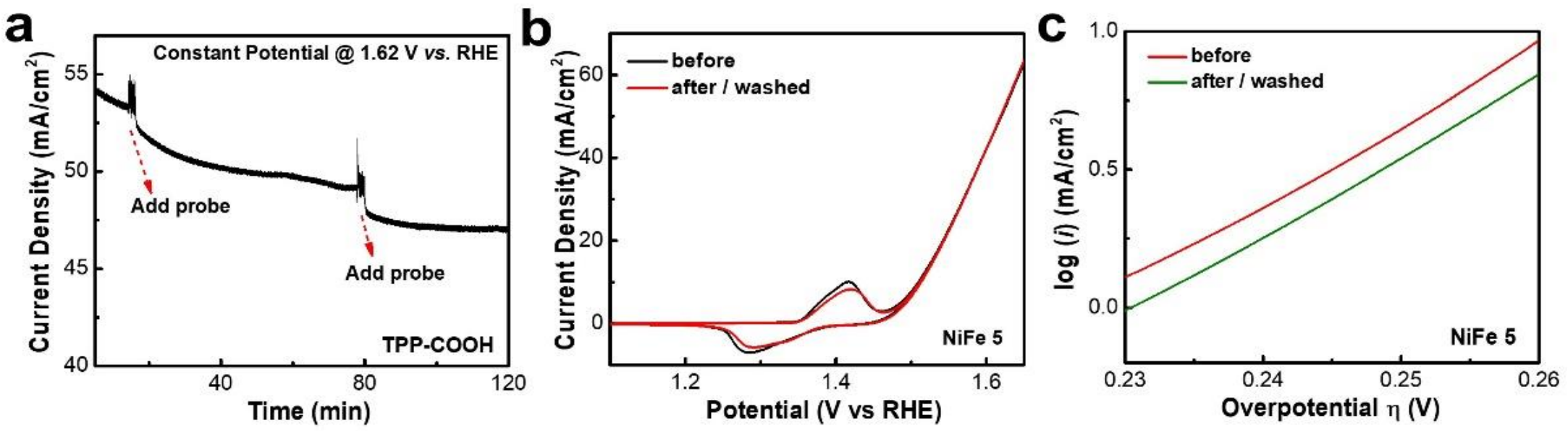

**Figure S18**. a) Constant potential electrolysis i-t curve of NiFe 5 in 1 M KOH at a constant potential of 1.62 V vs RHE with intermittent addition of 1 mM TPP-COOH. b-c) Cyclic voltammetry curves (b) and Tafel plots of the NiFe 5 catalyst in 1 M KOH before and after electrolysis with TPP-COOH. The electrodes were repetitively washed with water after the electrolysis. The scan rate is 5 mV/s for CV measurement and 0.5 mV/s for Tafel plot measurement. The resistance is 2.2Ω.

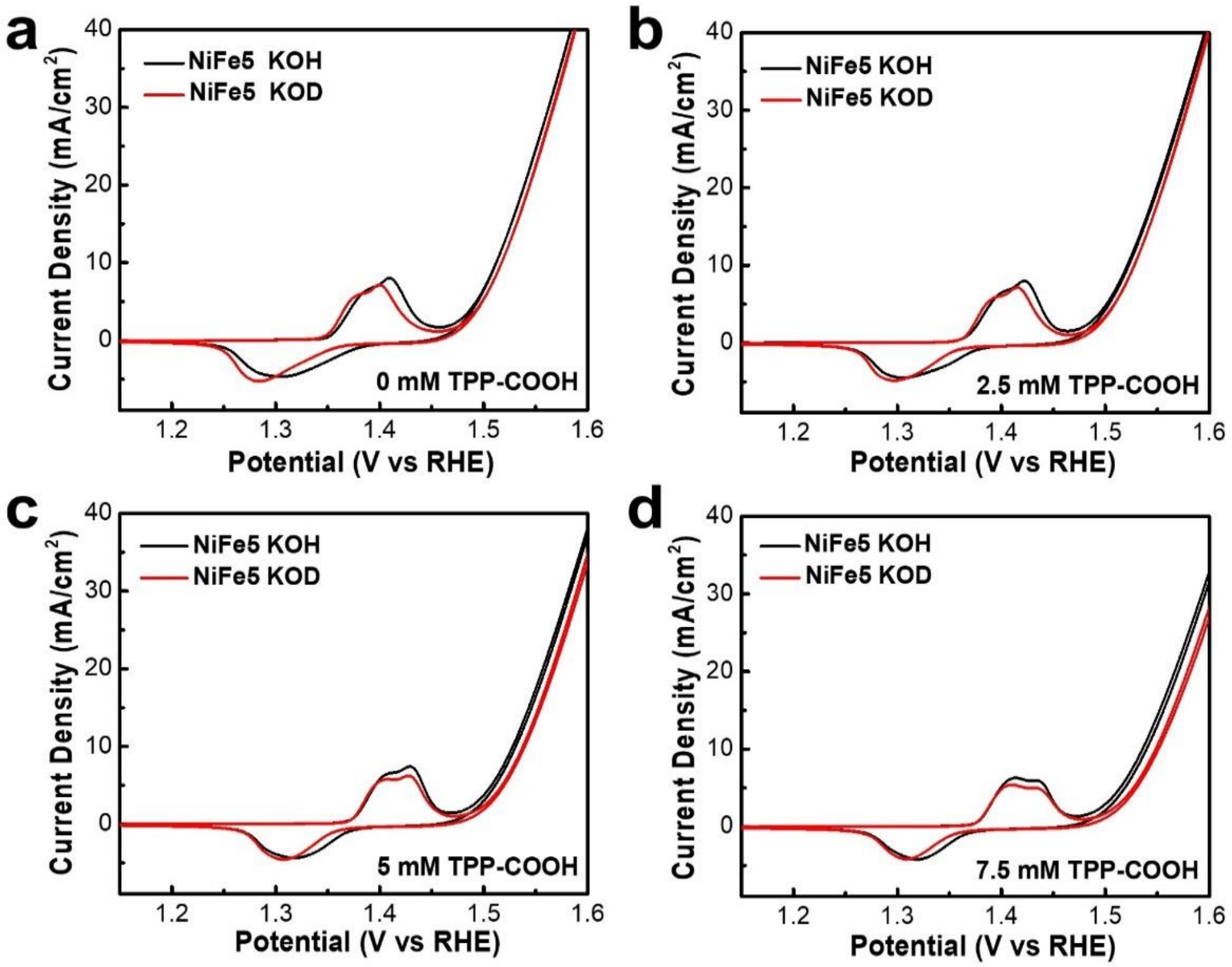

**Figure S19**. a-d) Cyclic voltammetry curves of the NiFe 5 catalyst in 1 M KOH/$H_2O$ and 1 M KOD/$D_2O$ electrolytes with different concentrations of TPP-COOH. The potential was corrected by the measured pH according to Nernst equation. The scan rate is 5 mV/s and the resistance is 1.8 ohm.

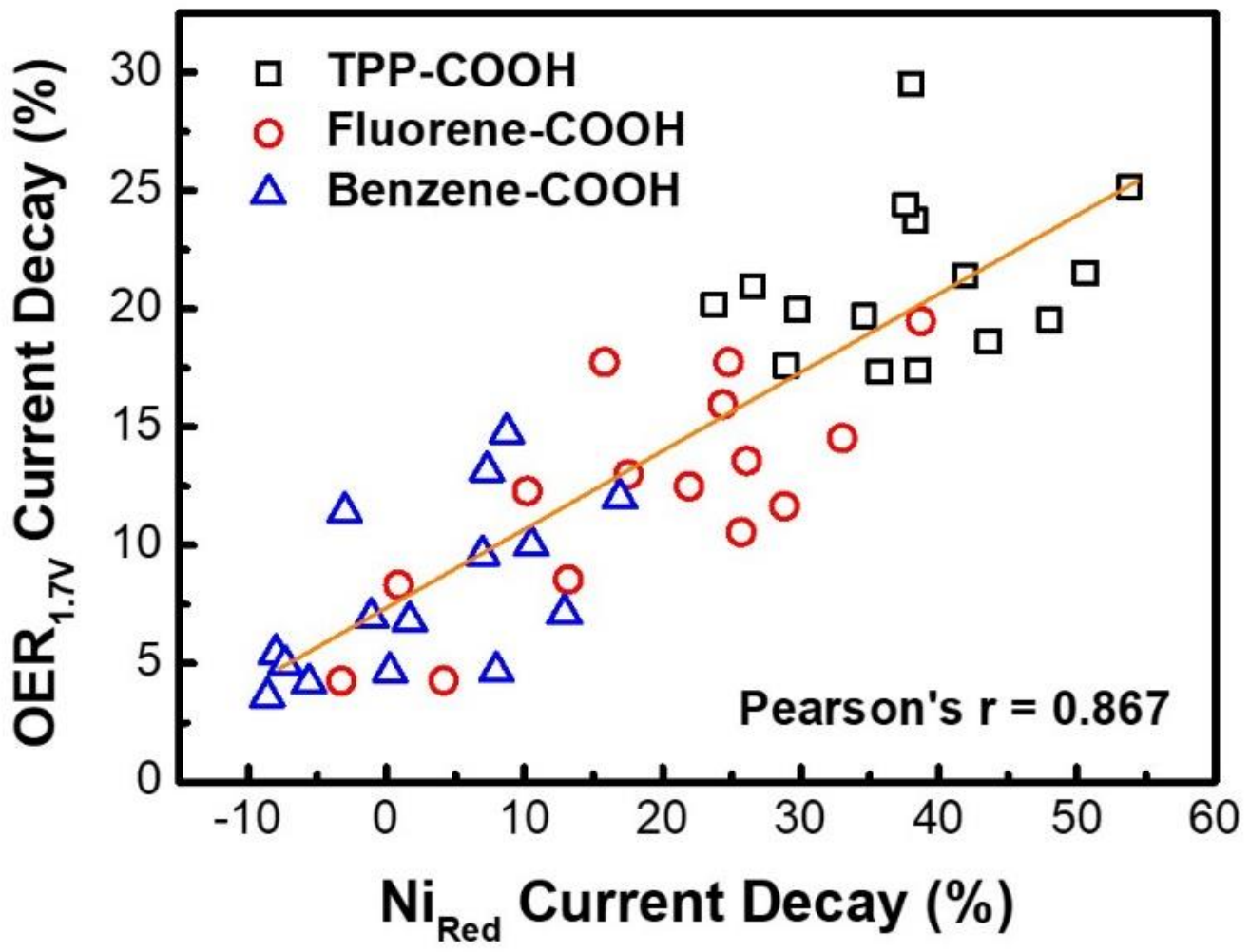

**Figure S20**. The correlation between the OER current and Ni reduction current decay upon the addition of reactive probes. The Pearson coefficient is calculated to be 0.867.

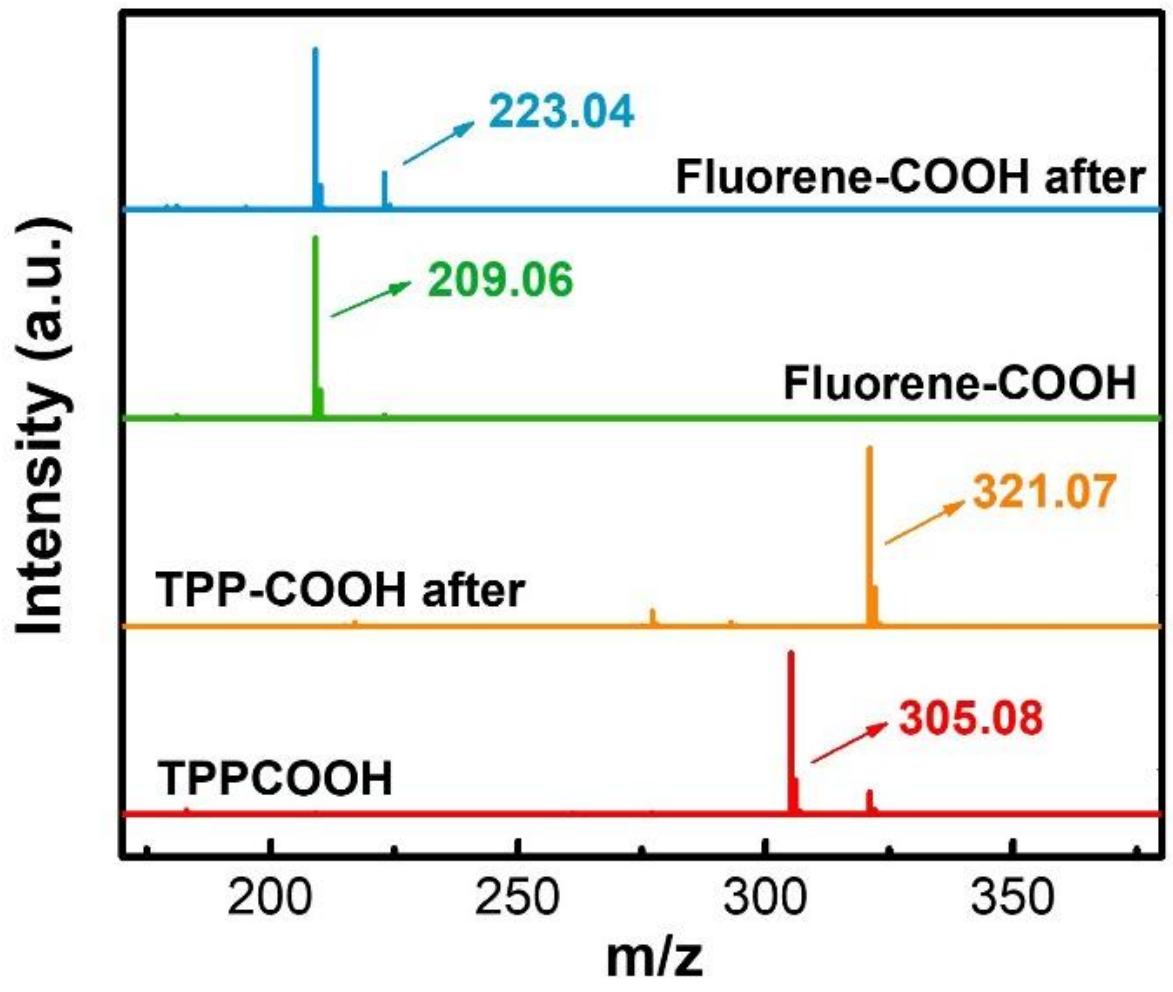

**Figure S21**. The mass spectra of the reactive probes of TPP-COOH and Fluorene-COOH before and after electrolysis

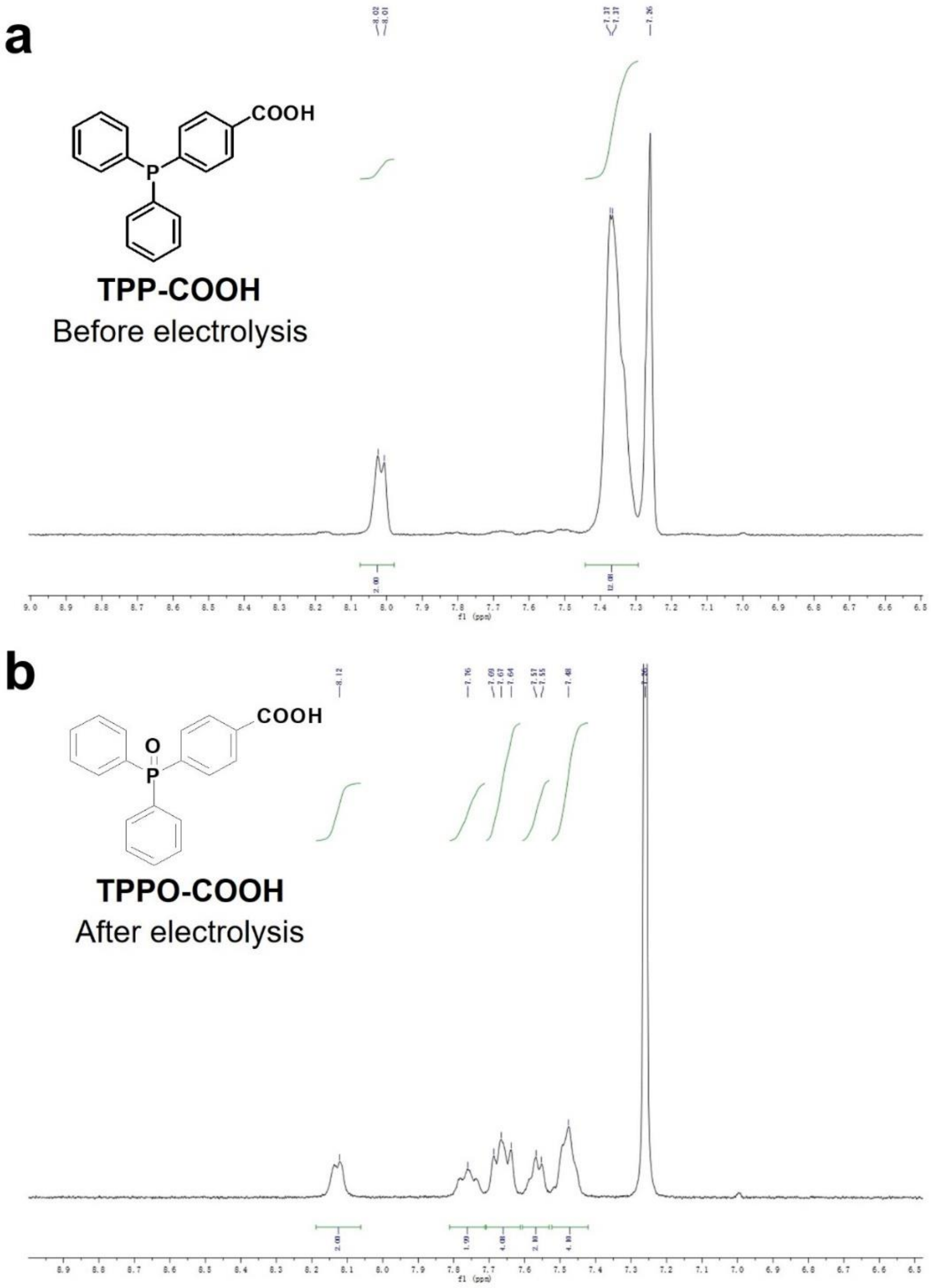

**Figure S22**. a) the NMR spectrum of TPP-COOH before electrolysis. b) the NMR spectrum of TPP-COOH after electrolysis, corresponding to TPPO-COOH.

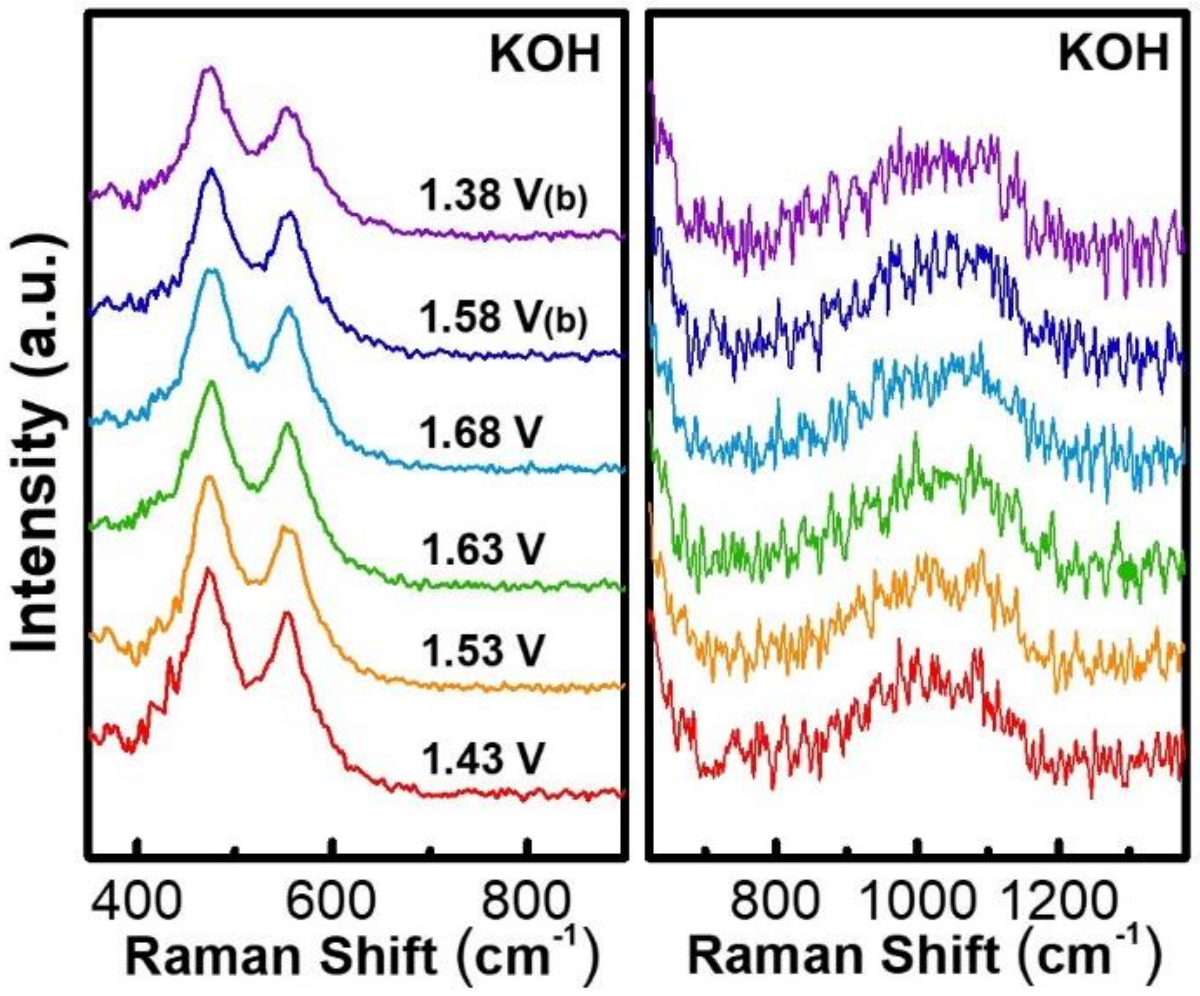

**Figure S23**. The Raman spectra of the NiFe 5 catalyst in 1 M KOH under different applied potentials

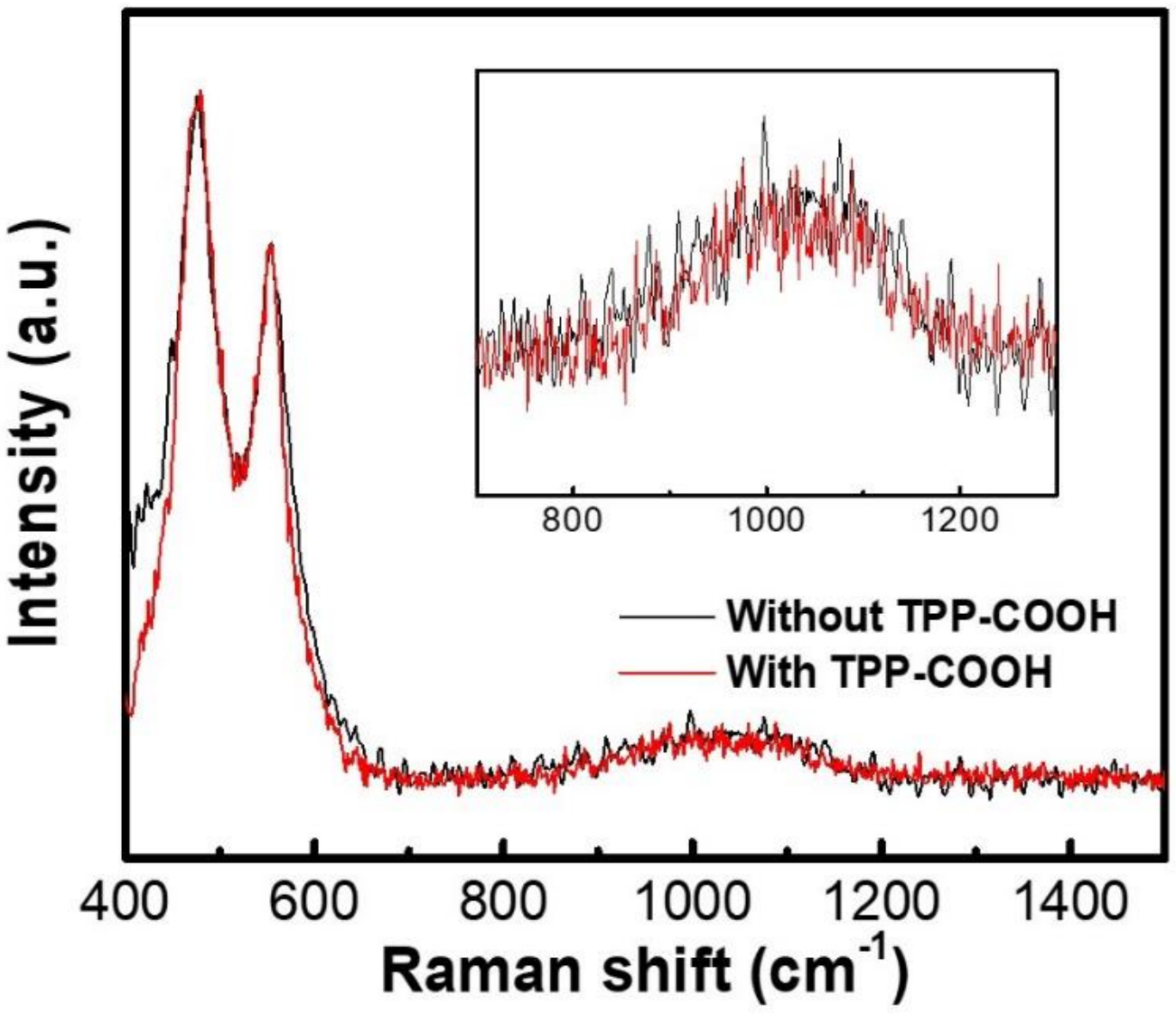

**Figure S24**. a) Normalized Raman spectrum of NiFe 5 catalyst in 1 M KOH with and without TPP-COOH under the potential of 1.59 V vs RHE, showing no drastic differences. The inset shows the magnified region of 800-1200 cm$^{-1}$.

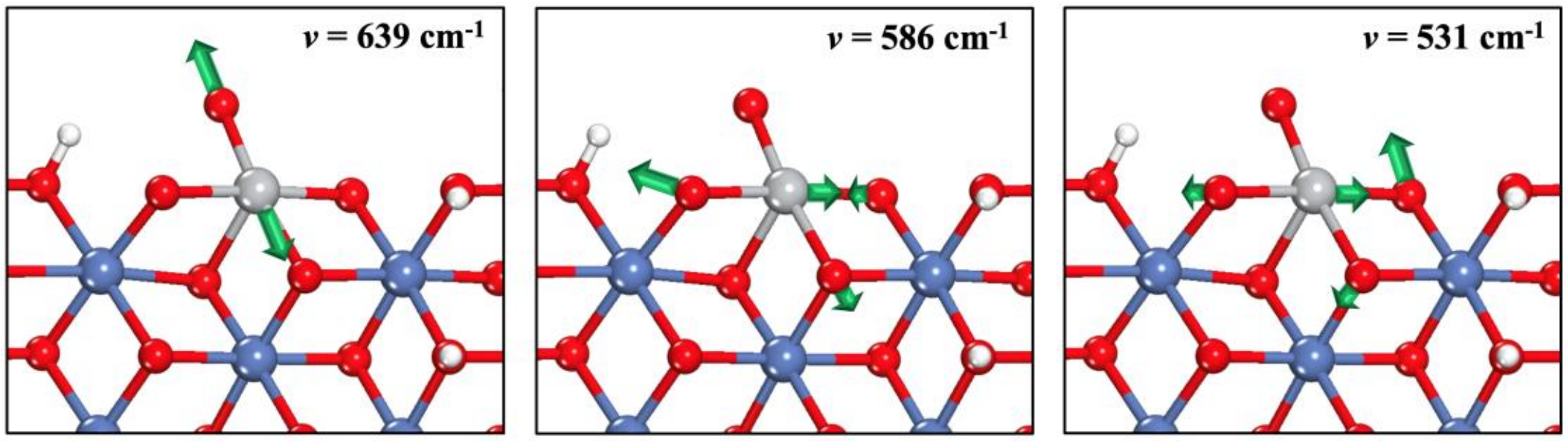

**Figure S25.** Modes of vibration for *$FeO(\mu_2\text{-}O)_2$, at frequencies of 639, 586, and 531 $cm^{-1}$. The atomic movement vectors are labeled using the green arrows. Blue balls: Ni; gray balls: Fe; red balls: O; white balls: H.

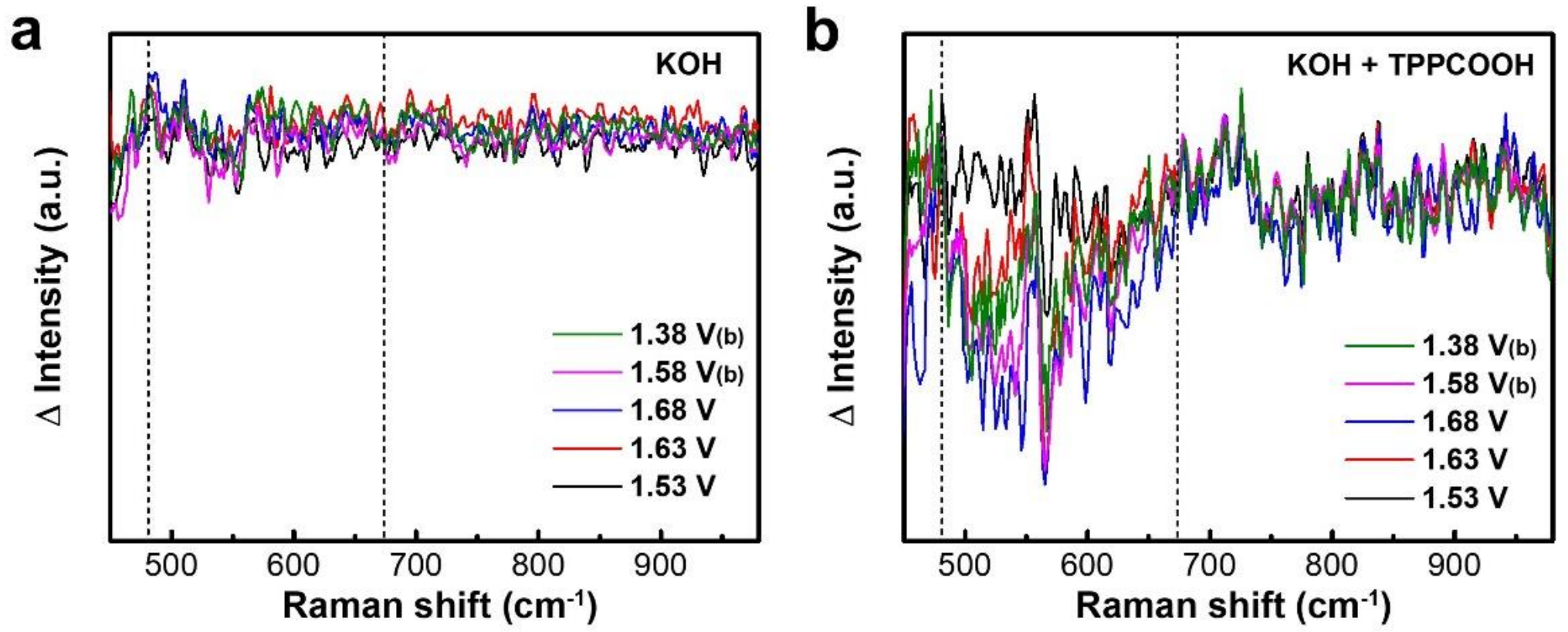

**Figure S27.** The difference in normalized Raman spectra of the NiFe 5 catalyst in 1 M KOH under different applied potentials in a) KOH and b) KOH + TPP-COOH. The spectrum measured under 1.43 V vs. RHE was utilized as the standard. The dashed line outlines the Raman shift region of the largest differences. Part of the >600 $cm^{-1}$ region is included and may involve the Fe=O intermediates.

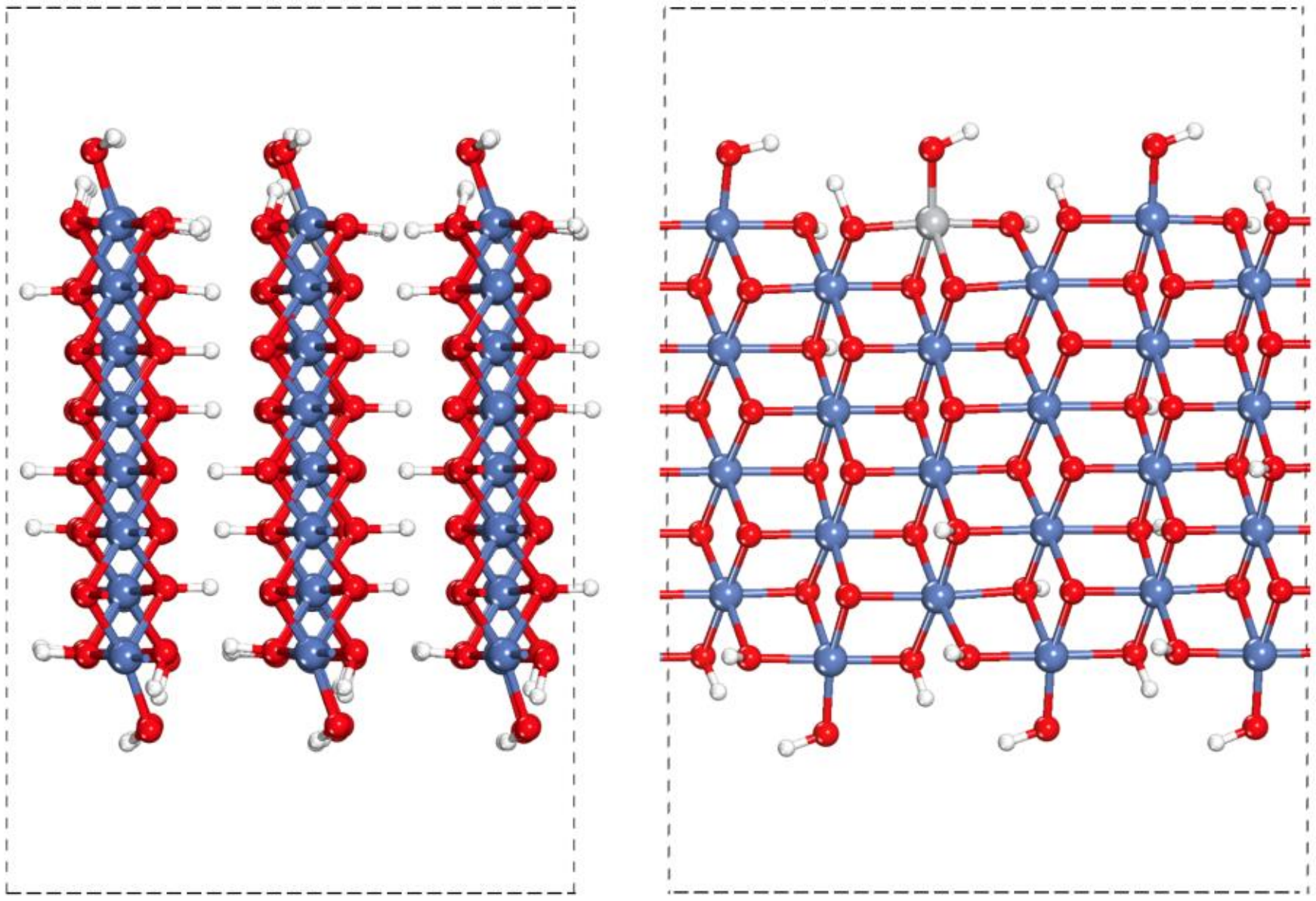

**Figure S27.** Atomic structures of γ-NiOOH $(\bar{1}2\bar{1}0)$ surfaces viewed from (a) γ-NiOOH $[1000]$ and (b) $[0001]$ directions. The dashed lines represent the boundary of the surface supercell. Blue balls: Ni; gray balls: Fe; red balls: O; white balls: H.

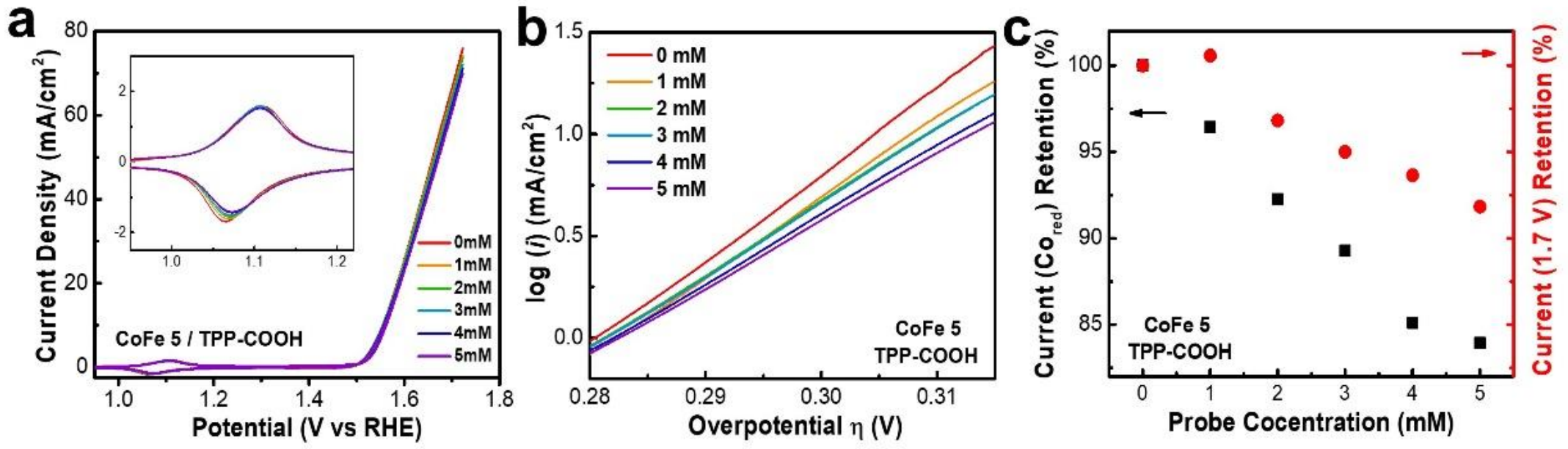

**Figure S28**. a) Cyclic voltammetry curves of CoFe 5 catalysts in 1 M KOH under the titration of 4-diphenylphosphino-benzoic acid (TPP-COOH). The scan rate is 5 mV/s and the resistance is 2.1 Ω. The inset shows the corresponding Co redox feature. b) Tafel plot of CoFe 5 catalyst in 1 M KOH under the titration of 4-diphenylphosphino-benzoic acid (TPP-COOH). The scan rate is 0.5 mV/s. c) The trend of $Co^{3+}$ reduction peak current and OER current over different TPP-COOH concentrations.

**Table S1.** The resistance of the electrochemical cell with NiFe 5 catalyst under different probes.

| Sample | Probe | | | | | | |
|---|---|---|---|---|---|---|---|
| | TPP-COOH | Styrene-COOH | Fluorene-COOH | Toluene-COOH | Benzene-COOH | Phenol-COOH | Thioanisole-COOH |
| | R/Ω | | | | | | |
| Ni | 2.0 | 2.3 | 2.5 | 2.4 | 2.4 | 2.4 | 2.1 |
| NiFe 1 | 1.9 | 2.2 | 2.6 | 2.0 | 2.3 | 2.5 | 2.0 |
| NiFe 5 | 1.8 | 2.3 | 2.5 | 2.2 | 2.4 | 1.9 | 2.1 |
| NiFe 10 | 2.1 | 2.1 | 2.5 | 2.1 | 2.1 | 2.3 | 2.2 |
| NiFe 20 | 2.0 | 2.1 | 2.3 | 1.9 | 2.0 | 2.0 | 2.1 |
| NiFe 50 | 2.0 | 2.4 | 2.2 | 2.4 | 2.6 | 2.3 | 2.2 |

## References

1. Kresse, G.; Furthmüller, J., Efficient iterative schemes for \textit{ab initio} total-energy calculations using a plane-wave basis set. *Phys. Rev. B* **1996,** *54* (16), 11169-11186.

2. Perdew, J. P.; Burke, K.; Ernzerhof, M., Generalized Gradient Approximation Made Simple. *Phys. Rev. Lett.* **1996,** *77* (18), 3865-3868.

3. Anisimov, V. I.; Zaanen, J.; Andersen, O. K., Band theory and Mott insulators: Hubbard U instead of Stoner I. *Phys. Rev. B* **1991,** *44* (3), 943-954.

4. Cococcioni, M.; De Gironcoli, S., Linear Response Approach to the Calculation of the Effective Interaction Parameters in the LDA+ U Method. *Phys. Rev. B* **2005,** *71* (3), 035105.

5. García-Mota, M.; Bajdich, M.; Viswanathan, V.; Vojvodic, A.; Bell, A. T.; Nørskov, J. K., Importance of Correlation in Determining Electrocatalytic Oxygen Evolution Activity on Cobalt Oxides. *J. Phys. Chem. C* **2012,** *116* (39), 21077-21082.

6. Li, Y.-F.; Selloni, A., Mechanism and Activity of Water Oxidation on Selected Surfaces of Pure and Fe-Doped NiOx. *ACS Catal.* **2014,** *4* (4), 1148-1153.

7. Martirez, J. M. P.; Carter, E. A., Unraveling Oxygen Evolution on Iron-Doped β-Nickel Oxyhydroxide: The Key Role of Highly Active Molecular-like Sites. *J. Am. Chem. Soc.* **2019,** *141* (1), 693-705.

8. Zaffran, J.; Caspary Toroker, M., Benchmarking Density Functional Theory Based Methods To Model NiOOH Material Properties: Hubbard and van der Waals Corrections vs Hybrid Functionals. *J. Chem. Theo. Comput.* **2016,** *12* (8), 3807-3812.

9. Li, Y.-F.; Selloni, A., Mosaic Texture and Double c-Axis Periodicity of β-NiOOH: Insights from First-Principles and Genetic Algorithm Calculations. *The Journal of Physical Chemistry Letters* **2014,** *5* (22), 3981-3985.

10. Henkelman, G.; Jónsson, H., A dimer method for finding saddle points on high dimensional potential surfaces using only first derivatives. *J. Chem. Phys.* **1999,** *111* (15), 7010-7022.

11. Heyden, A.; Bell, A. T.; Keil, F. J., Efficient methods for finding transition states in chemical reactions: Comparison of improved dimer method and partitioned rational function optimization method. *J. Chem. Phys.* **2005,** *123* (22), 224101.

12. Fattebert, J. L.; Gygi, F., Linear-scaling first-principles molecular dynamics with plane-waves accuracy. *Phys. Rev. B* **2006,** *73* (11), 115124.

13. Fang, Y. H.; Liu, Z. P., Mechanism and Tafel Lines of Electro-Oxidation of Water to Oxygen on $RuO_2$(110). *J. Am. Chem. Soc.* **2010,** *132* (51), 18214-18222.

14. Li, Y. F.; Liu, Z. P.; Liu, L.; Gao, W., Mechanism and Activity of Photocatalytic Oxygen Evolution on Titania Anatase in Aqueous Surroundings. *J. Am. Chem. Soc.* **2010,** *132* (37), 13008-13015.

15. Li, Y. F.; Liu, Z. P., Active Site Revealed for Water Oxidation on Electrochemically Induced $\delta$-$MnO_2$: Role of Spinel-to-Layer Phase Transition. *J. Am. Chem. Soc.* **2018,** *140* (5), 1783-1792.